\documentclass[journal,a4paper,twoside]{IEEEtran}

\usepackage{amsmath}
\usepackage[tight,footnotesize]{subfigure}

\usepackage[utf8]{inputenc} % allow utf-8 input
\usepackage[T1]{fontenc}    % use 8-bit T1 fonts
\usepackage{hyperref}       % hyperlinks
\usepackage{url}            % simple URL typesetting
\usepackage{xurl}
\usepackage{booktabs}       % professional-quality tables
\usepackage{amsfonts}       % blackboard math symbols
\usepackage{nicefrac}       % compact symbols for 1/2, etc.
\usepackage{microtype}      % microtypography
\usepackage{url}
\usepackage{cite} 
\usepackage{cleveref}

\crefname{equation}{Eq.}{Eqs.}
\crefname{pluralequation}{Eqs.}{Eqs.}

\crefname{algorithm}{Algorithm}{Algorithm}

\crefname{figure}{Fig.}{Figs.}
\crefname{pluralfigure}{Figs.}{Figs.}

\crefname{section}{Sect.}{Sects.}
\crefname{pluralsection}{Sects.}{Sects.}

\crefname{table}{Table}{Table}
\crefname{pluraltable}{Tables}{Tables}

\crefname{definition}{Def.}{Def.}
\crefname{pluraldefinition}{Defs.}{Defs.}

\crefname{theorem}{Theorem}{Theorems}
\crefname{pluraltheorem}{Theorems}{Theorems}

\crefname{lemma}{Lemma}{Lemmas}
\crefname{plurallemma}{Lemmas}{Lemmas}

\crefname{example}{Example}{Example}
\crefname{pluralexample}{Examples}{Examples}

\crefname{problem}{Problem}{Problem}
\crefname{pluralproblem}{Problems}{Problems}

\crefname{assumption}{Assumption}{Assumption}
\crefname{pluralassumption}{Assumptions}{Assumptions}

\crefname{remark}{Remark}{Remark}
\crefname{pluralremark}{Remarks}{Remarks}

\crefname{appendix}{Appendix}{Appendices}
\crefname{pluralappendix}{Appendices}{Appendices}

\usepackage{graphics}
\usepackage{tabularx}
\usepackage{multirow}
\usepackage{booktabs}
\usepackage{graphicx}
\usepackage{epsfig}
\usepackage{mathrsfs}
\usepackage{color}
\usepackage{amssymb}
\usepackage{amsmath}
\usepackage{amsthm}
\usepackage{algorithm}
\usepackage{algpseudocode}
\usepackage{mathtools}

\usepackage{subfigure}
\usepackage{euscript}
\usepackage{bm}
\usepackage{verbatim}
\usepackage[normalem]{ulem}
\usepackage{xcolor}
\usepackage{comment}
\usepackage{silence}

\usepackage{color}
\usepackage{epsfig} 
\graphicspath{{.}{./figs/}}
\DeclareGraphicsExtensions{.eps}

\def\bfn#1{\bm{\mathbf{#1}}}    % Bold em math mode (requires) 
\DeclareMathOperator*{\argmax}{arg\,max}

\def\req#1{(\ref{#1})}

\def\real{\mathbb{R}}

\def\req#1{(\ref{#1})}

\theoremstyle{thmstyleone}%
\theoremstyle{thmstyletwo}%

\theoremstyle{thmstylethree}%

\newcommand\changes[1]{\textcolor{black}{#1}}
 
\begin{document}
\def\abstractt{
New residential neighborhoods are often supplied with heat via district heating systems (DHS). Improving the energy efficiency of a DHS is critical for increasing sustainability and satisfying user requirements. In this paper, we present HELIOS, a dedicated artificial intelligence (AI) model designed specifically for modeling the heat load in DHS. HELIOS leverages a combination of established physical principles and expert knowledge, resulting in superior performance compared to existing state-of-the-art models. HELIOS is explainable, enabling enhanced accountability and traceability in its predictions. We evaluate HELIOS against ten state-of-the-art data-driven models in modeling the heat load in a DHS case study in the Netherlands. HELIOS emerges as the top-performing model while maintaining complete accountability. The applications of HELIOS extend beyond the present case study, potentially supporting the adoption of AI by DHS and contributing to sustainable energy management on a larger scale. 
}

\def\titulo{Integrating Expert and Physics Knowledge for Modeling Heat Load in District Heating Systems}

% FOR IEEE TEMPLATE
\title{\titulo}
\author{Francisco~Souza,  Thom~Badings, Geert~Postma, Jeroen~Jansen
\thanks{
All authors are with Radboud University, Nijmegen, The Netherlands.
This work was supported by NWO grant NWA.1160.18.238 (PrimaVera), and by the European Union’s Horizon 2020
research and innovation programme under grant agreement N. 952003 (AI-REGIO).  Emails: \footnotesize\{francisco.souza,~thom.badings,~g.postma,~jeroen.jansen\}@ru.nl \\
©2025 IEEE. Personal use of this material is permitted. Permission from IEEE must be
obtained for all other uses, in any current or future media, including
reprinting/republishing this material for advertising or promotional purposes, creating new
collective works, for resale or redistribution to servers or lists, or reuse of any copyrighted
component of this work in other works

}

}
\vspace{-.5cm}

\maketitle

\begin{abstract}
\abstractt
\end{abstract}
\begin{IEEEkeywords}
District heat systems, expert systems, integrating expert knowledge, physics-guided models, forecasting
%   \tb{Keywords not completed yet!}
\end{IEEEkeywords}

\IEEEpeerreviewmaketitle

%---------------------------------------------------------------------------
%              New Section
%---------------------------------------------------------------------------
% \vspace{-0.15cm}
\section{Introduction}
\label{sec:intro}
District heating systems (DHS) are piping networks for distributing heat from a centralized location to connected end-users~\cite{benonysson1995operational}.
Improving the energy efficiency of a DHS is critical for increasing operational sustainability and meeting rising demands. 
However, optimizing DHS is difficult due to a lack of historical data and effective optimization methods~\cite{Mbiydzenyuy2022}.
DHS digitalization aims to solve these challenges by developing optimization models based on artificial intelligence (AI) that combine the use of data with forecasting models and control technologies.
Thus, effective DHS digitalization depends on accurately modeling the head load of DHS~\cite{Zdravkovic2022}, which is the main problem we consider in this paper.

Approaches for modeling the heat load in DHS can generally be distinguished between physics-based (leveraging physics and behavioral models) and data-driven (leveraging AI models).
Physics-based models typically use heat balances to define prediction models for the heat load~\cite{Nielsen2006}.
The behavioral customer profile, often related to hot water consumption, is approximated indirectly from occupancy features (e.g.,\ working hours, day of the week, etc.) of a parameterized model~\cite{Lund2020}. 
Physics-based models can often outperform data-driven models~\cite{Willcox2021}.
{
Model predictive control (MPC) is a popular approach to process optimization using physics-based models~\cite{aoun2019modelling,verrilli2016model}.
However, classical MPC requires a precise definition of the dynamics model of the system.
As a result, expert knowledge has to be captured directly in the dynamics model (e.g., by setting parameters and coefficients in the dynamics), which can be a complex task in practice.
Nevertheless, integrating physics and expert knowledge is key to optimizing DHS operations.
}

Data-driven AI models, on the other hand, assume either a specific functional form for the model~\cite{Dotzauer2002} or estimate the functional form from data, e.g.,\ using tree-based models or deep learning (DL)~\cite{Potocnik2015,Idowu2016,Dahl2017,Suryanarayana2018,Xue2019,Bunning2020,Bergsteinsson2021,Castellini2021}. 
One of the first data-driven approaches is presented by \cite{Dotzauer2002}, who use a piecewise model for modeling the heat load from weather data and occupational features for modeling the consumer behavior profile.
Often, these occupational features are indirectly approximated as time series. 
The papers~\cite{Potocnik2015, Dahl2017,Idowu2016,Suryanarayana2018,Xue2019,Castellini2021} investigate several AI models for modeling the heat load, using the ambient temperature and occupancy variables as model input. 
The authors conclude that the ambient temperature, time, and day of the week are the most important factors in predicting heat load.
AI models are used for multi-step and online predictions by~\cite{Xue2019,Bunning2020}; however, patterns and user behavior in DHS are reasonably constant over time, so such an approach may not be well-suited for DHS modeling.
Modeling the heat load via DL models, e.g., recurrent neural networks (RNNs), long short-term memory (LSTM), and temporal hierarchy models, was done by~\cite{Kato2008,Lin2020,Song2020,Lu2021,Bergsteinsson2021,Zdravkovic2022}, reporting improved performance over shallow neural networks and classical machine learning models.
For a recent survey of data-driven DHS modeling, we refer to~\cite{Ntakolia2022}. 
%\vspace{-20pt}

However, most data-driven heat load modeling methods based on AI lack transparency and trustworthiness~\cite{Ntakolia2022}.
These data-driven methods often neglect the information on process operation from process experts in DHS modeling, i.e., they do not sufficiently incorporate the available expert knowledge.
Moreover, purely data-driven methods suffer from limited explainability, resulting in a lack of acceptance by stakeholders.
Since explainability is crucial for stakeholder support, deploying AI-based solutions in risk or engineering systems is thus hampered~\cite{Rudin2022}. 
Overall, few existing approaches use physical information to model the heat load in DHS, and none integrate expert knowledge into the final model. 

% \vspace{-0.35cm}
\subsection*{Our Approach and Contributions}
\changes{
In this paper, we propose HELIOS,\footnote{The name is inspired by Helios, the god of the Sun in Greek mythology.} a novel expert-physics-based model to predict the heat load demand in DHS. 
As a novel feature compared to existing methods, we decompose the heat load into three meaningful physical components: (1)~space heating, (2)~hot water consumption, and (3)~transmission losses. 
For each of these three components, we combine physics-based models with statistical models designed to reflect the specificities of each component (e.g., reflecting heating profiles and customer behavior).
As a unique feature, we leverage the existing method from~\cite{Souza2022} to integrate expert knowledge into the learning stage, using the specificities of the DHS to enhance the physics-based models.
Existing DHS prediction methods lack such capability for integrating expert knowledge.
}

\changes{
In summary, our contributions are twofold: (1) We propose a novel model for predicting the head load demand in DHS, which combines established physical principles with expert knowledge; (2) We benchmark our method against 14 statistical models widely employed for modeling the heat load in DHS.
Notably, our HELIOS model shows better performance than all alternative approaches while at the same time providing meaningful insights on the heat load profile of the DHS.}
% \vspace{-0.2cm}

%---------------------------------------------------------------------------
%              New Section
%---------------------------------------------------------------------------

\section{Overview of the HELIOS Model}
\label{sec:Helios}

In this section, we present the HELIOS model from a top-down perspective.
First, we provide a general overview of HELIOS in \cref{sec:Helios:model}.
In \cref{sec:Helios:ExpertModels}, we then introduce so-called contextual mixture of expert models, which we use to capture expert knowledge in HELIOS.
In \cref{sec:Helios:space_heating,sec:Helios:hot_water,sec:Helios:heat_loss}, we discuss the individual components of HELIOS in detail.
Finally, we provide a complete overview of the HELIOS model and describe how we learn its parameters in \cref{ss.learning}.
% \vspace{-0.3cm}

%---------------------------------------------------------------------------
%              New Subsection
%---------------------------------------------------------------------------
\subsubsection*{Notation}
\changes{Throughout the paper, vectors are represented by boldface lowercase letters, e.g.,\ $\bfn{a}=[a_1,\ldots,a_d]^\top$. 
Random variables are denoted by capital letters, and a realization of the random variable $A$ at sampling time $k$ is written as $a(k)$ (and as $\bfn{a}(k)$ in case of a vector).}

% \vspace{-0.3cm}

%---------------------------------------------------------------------------
%              New Subsection
%---------------------------------------------------------------------------
\subsection{Causal Graph Model of HELIOS} 
\label{sec:Helios:model}

\begin{figure}[!t]
\centering
\includegraphics[width=1.\columnwidth]{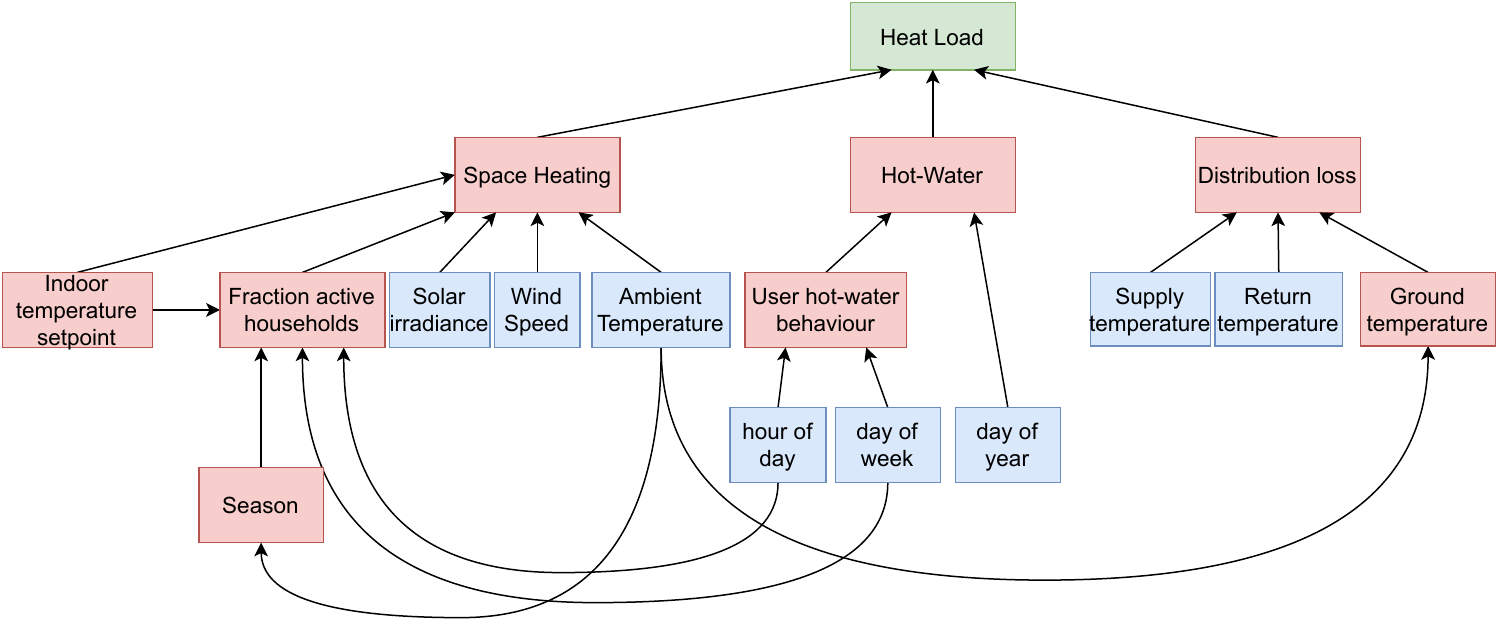}
\caption{Causal representation of the factors of influence in the heat load.}
\label{fig.helios_causal}
\vspace{-0.35cm}
\end{figure}

The architecture of the HELIOS model is built upon the causal graph model shown in Fig.\ \ref{fig.helios_causal}, representing the factors influencing the heat load.
Using such a causal graph provides a principled approach for AI to leverage expert knowledge \cite{Teshima2021}. 
We derive this causal model from existing DHS knowledge gathered from literature and stakeholder interviews. 

The heat load is defined as \emph{``the heat supply needed to satisfy the customer's heat demand.''}~\cite{Frederiksen2013} 
Doing so, we decompose the heat load $Q(k)$ at time $k$ (the top node in \cref{fig.helios_causal}) into three unmeasured variables: the space heating $Q_{\text{space}}(k)$, the domestic hot water consumption $Q_{\text{hot-water}}(k)$, and the transmission piping loss $Q_{\text{loss}}(k)$.
Thus, we obtain the heat balance
%The heat load $Q$ is computed as the sum of these contributions:
%
\begin{align}
    Q(k) = Q_{\text{space}}(k) + Q_{\text{hot-water}}(k) + Q_{\text{loss}}(k).
\label{equ.fq}
\end{align}
Space heating is \emph{``the heat necessary to create a comfortable indoor climate,''} hot water consumption is \emph{``the heating required to warm up water in the tap points according to the consumers needs,''}  and the piping loss is \emph{``the heat lost along the pipes during transmission.''}
Decomposing $Q$ in \cref{equ.fq} improves the accountability of these three factors to the total heat load, which in turn improves the explainability of predictions.

\relax

{
\subsection{Contextual Mixture of Experts Models}
\label{sec:Helios:ExpertModels}

We incorporate expert knowledge in HELIOS using so-called \emph{contextual mixture of experts (cMoE) models}~\cite{Souza2022}. 
Intuitively, the cMoE model integrates a collection $\CMcal{C}$ of \emph{experts}, denoted as $\{f_c\}_{c \in \CMcal{C}}$, along with a corresponding set of \emph{gating mechanisms}, $\{g_c\}_{c \in \CMcal{C}}$. 
The output of the cMoE model for an input $\bfn{x}(k)$ is 
\begin{equation}
f(\bfn{x}(k); \Theta ) = \sum_{c \in \mathcal{C}} g_c(\bfn{x}(k); \theta_{gc}) \cdot f_c(\bfn{x}(k); \theta_{ec}),
\end{equation}
where $\Theta=\{\theta_{gc}, \theta_{ec}\}_{c \in \CMcal{C}}$ is the set of gates $\theta_{gc}$ and experts $\theta_{ec}$ parameters. 
Each item $c \in \CMcal{C}$ is called a \emph{context} and models a significant attribute or mode of the system.
In practice, contexts are often derived from domain-specific knowledge. % and can thus be identified by domain experts. 

The objective is to optimize the parameters $\Theta$ of the function $f(\Theta) \colon X \rightarrow Y$ given a dataset $\mathcal{D}$.
This dataset is defined as $\mathcal{D} = \{(\bfn{x}(k), y(k), \bfn{\pi}(k) )\}_{k=1}^N$ and consists of $N$ samples, with inputs $\bfn{x}(k) \in \real^D$, $D \in \mathbb{N}$, target (or output) variables $y(k) \in \real$, and context variables $\bfn{\pi}(k) \in \real^{|\CMcal{C}|}$.
The context variables are defined from \emph{possibility distributions}~\cite{DUBOIS2014}, which express the belief that sample $k$ is part of context $c$.
We denote the context variable for context $c$ and sample $k$ as $\pi_c(\mathbf{x}(k)) \in [0,1]$, or $\pi_{c}(k)$ in short. 
Thus, $\pi_{c}(k)$ reflects the \emph{degree of possibility} that sample $k$ is part of the context $c$. 
If $\pi_{c}(k) = 1$, then sample $k$ is fully \emph{compatible} (or consistent) with context $c$, denoting a complete possibility of belonging to that context. 
Conversely, $\pi_{c}(k) = 0$ means that sample $k$ is not part of context $c$, i.e., the sample cannot belong to that context. For other values $\pi_{c}(k) \in (0,1)$, the sample $k$ has a partial possibility of belonging to context $c$. 
This range represents degrees of uncertainty or partial belief regarding the sample's association with the context. 
In this paper, we employ the $\alpha$-Certainty possibility distribution \cite{DUBOIS2014} to represent contexts.

\paragraph*{$\alpha-$certain possibility distribution} 
We use the $\alpha$-Certain possibility distribution from~\cite{DUBOIS2014}, which incorporates a certainty factor $\alpha \in [0,1]$ to quantify the level of confidence associated with the knowledge about the context $c \in A$ (intuitively, ``$c$ is certain to a degree $\alpha$''), and $A$ is the support of context $c$ in the sample space.
As such, we obtain the model
\begin{equation}
    \pi_{c}(k) =
    \begin{cases}
      1 \enskip& \text{if $k \in  A$},\\
      1-\alpha & \text{otherwise}.\\
    \end{cases}
\end{equation}
If $\alpha=1$, then $\pi_{c}(k)$ represents the characteristic function of context $c$.
On the other hand, for $\alpha=0$, $\pi_{c}(k)$ represents the total ignorance about context $c$. 
%For further details on cMoE we refer to~\cite{Souza2022}.

}

{
\section{Modeling Space Heating in HELIOS}
\label{sec:Helios:space_heating}
}

Space heating ensures a comfortable indoor climate when the ambient temperature is below the desired indoor temperature \cite{Frederiksen2013}.
We model the space heating using a physics-based model of the buildings connected to the DHS.
Concretely, the space heating $Q_{\text{space}}(k)$ at time $k$ in \cref{equ.fq} is defined as the total heat load in $B \in \mathbb{N}$ buildings connected to the DHS:
\begin{equation}
    {Q}_{\text{space}}(k) 
    = \sum_{b=1}^B {Q}_{\text{space},b}(k),
\label{equ.space_energy_balance_all_buildings}
\end{equation}
where the space heating ${Q}_{\text{space},b}(k)$ for building $b$ depends on the setpoint temperature $T_\text{set}$ and the building characteristics.
We present the space heating model in three steps.
First, in \cref{subsec:temperature_setpoint}, we describe a model for the desired indoor (i.e., the \emph{setpoint}) temperature of buildings.
Second, in \cref{subsec:active_households}, we introduce the so-called \emph{fraction of active households}, as an abstraction for aggregating buildings.
Third, in \cref{sec:thermodynamics_single}, we describe a detailed thermodynamic model for individual buildings.
We finally combine these ingredients in \cref{subsec:aggregated_space_heating} to obtain the final space heating model for HELIOS.

\subsection{Indoor Temperature Setpoint Model} 
\label{subsec:temperature_setpoint}
The temperature setpoint $T_\text{set}$ regulates the indoor temperature $T_i$.
Ideally, the actual indoor temperature corresponds to the setpoint, i.e., $T_i = T_{\text{set}}$. 
The temperature setpoint has two different target temperatures: 1) the \emph{user comfort temperature}, which varies over time and depends on the occupancy profiles and status of the building, and 2) the \emph{setback temperature}, which is the setpoint when the building is not occupied~\cite{Santin2017}.
We employ the contextual mixture of experts model from~\cite{Souza2022} to model how the temperature setpoint varies with the hour of the day and the type of the day.
This results in the model
\begin{align}
    \widehat{T}_{\text{set}}(k) = \sum_{c \in \CMcal{C}_{T_\text{set}}} P^{(T_\text{set})}_{c,d(k)}( h(k))\zeta_{c,d(k)},
    \label{equ.Ts}
\end{align}
where $\widehat{T}_{\text{set}}$ is the approximated temperature setpoint for type of day $d$, and $\CMcal{C}_{T_\text{set}}$ is a set of model \emph{contexts} that capture knowledge of the setpoint. We consider two types of days: weekdays (1) and weekends and holidays (2), so $d(k) \in \{1,2\}$ for all $k$.
To match the setting with two different target temperatures, we define the set of contexts as $\CMcal{C}_{T_\text{set}}=\{\text{setback},\text{comfort}\}$
(we discuss the representation and integration of contextual information in Section~\ref{ss.learning}).
Every context $c$ and type of day $d$ has a dedicated parameter $\zeta_{c,d(k)}$.
Function $P^{(T_s)}_{c,d(k)}(h(k))$ is the probability that context $c \in \CMcal{C}_{T_\text{set}}$ holds at the hour $h(k)$ for the type of day $d(k)$, which we define as the softmax function
\begin{align}
    P^{(T_\text{set})}_{c,d(k)}(h(k)) = \exp(z_{c,d(k)})/\sum_{c \in \CMcal{C}_{T_\text{set}}} \exp(z_{c,d(k)}),
    \label{equ.PTs}
\end{align}
with $z_{c,d(k)}=\bfn{r}(h(k))\bfn{v}^T_{c,d(k)}$ and~$\bfn{z}_{d(k)}=[z_{1,{d}(k)},\ldots,z_{|\CMcal{C}_{T_\text{set}}|,d(k)}]$. Vector function $\bfn{r}(h(k)) \in \real^{2N}$ maps the hour of the day $h(k)$ to a vector given by a Fourier feature representation {$\bfn{r}(h(k))= \left[ \hat{r}_{1}(h(k)),\bar{r}_{1}(h(k)),\ldots,\hat{r}_{P}(h(k)),\bar{r}_{P}(h(k)) \right]^T$, where
\begin{align}
    {\hat{r}}_p(h(k)) = \sin\left(\frac{2\pi p h(k)}{24}\right) \,\,\,\,\,
    {\bar{r}}_p(h(k)) = \cos\left(\frac{2\pi p h(k)}{24}\right).
    \label{equ.hourToVec}
\end{align}
and $2P$ is the number of Fourier components.}
Overall, the set of unknown parameters in the temperature setpoint model (which we will learn in \cref{ss.learning})) is
\begin{equation}
\begin{split}
\Theta_{T_\text{set}}=\{\zeta_{c,d},\bfn{z}_{c,d} \mid  (c,d) \in \CMcal{C}_{T_\text{set}} \times \{1,2\} \}.
\end{split}
\end{equation}
The parameters in $\Theta_{T_\text{set}}$ are those to be learned in \cref{ss.learning}

%---------------------------------------------------------------------------
%              New Section
%---------------------------------------------------------------------------
\subsection{Fraction of Active Households} 
\label{subsec:active_households}
{
Modeling all buildings connected to the DHS individually is infeasible.
Thus, we define $A\in [0,1]$ as the fraction of \emph{active households}, which are those buildings whose temperature $T_i$ has not reached the setpoint, i.e., $T_i<T_\text{set}$. 
}
We assume that $A$ is a function of the hour of the day $h$, type of day $d$ (we again consider $d \in \{1,2\}$ for weekdays and weekends, respectively), and season $S$ (e.g., in summer it often holds that $T_i>T_\text{set}$).
To remove short-term fluctuations and emphasize longer-term seasonality, we model the season $S$ using a moving average from the ambient temperature measurement:
\begin{align}
    S(k) = \frac{1}{M} \sum_{m=1}^M T_a(k-m),
\end{align}
with $M$ the moving average length. 
It is desirable to select a large $M$ to express the seasonality pattern, e.g., in hourly measurements we set $M=24\times90$ ($24$ hours times $90$ days).
We use the season variable to compute the approximate fraction of active households $\widehat{A}$ as the probability that $T_i(k)<T_\text{set}(k)$:
\begin{align}
    \widehat{A}(k)=\alpha_s(k)\alpha_{i}(k),
    \label{equ.Ak}
\end{align}
where $\alpha_s(k)=P\left(T_i(k)<T_\text{set}(k) \big | S(k)\right)$ and $\alpha_{i}(k)=P\left(T_i(k)<T_\text{set}(k) \big | h(k), d\right)$ are the season and time influence, respectively.
We model both terms by a contextual mixture of experts model of the following form:%
\begin{align}
\alpha_s(k) &= \sum_{c \in \CMcal{C}_S}^{} P^{(A| S)}_c(S(k)) \eta_{c}
\label{equ.alpha_s}
\\
\alpha_{i}(k) &= \sum_{c\in \CMcal{C}_{T_\text{set}}}^{} P^{(A| T_\text{set})}_{c,d(k)}(h(k)) \mu_{c,d(k)},
\label{equ.alpha_i}
\end{align}
where $\CMcal{C}_S$ and $\CMcal{C}_{T_\text{set}}$ are the season and temperature setpoint context sets,
$P^{(A|S)}_c(S(k))$ and $P^{(A| T_s)}_{c,d(k)}(h(k))$ are the probabilities for each of these contexts, and $\eta_{c}$ and $\mu_{c,d(k)}$ are unit simplexes for the fraction of active households in context $c$ given the hour and type of the day.
For example, if we define $\CMcal{C}_{S}=\{\text{hot},\text{cold}\}$, we expect $\eta_\text{cold}>\eta_\text{hot}$ as colder seasons have higher space heating requirements. 
The probabilities in \cref{equ.alpha_s,equ.alpha_i} are defined as the following softmax functions: 
\begin{align}
    P^{(A| S)}_c(S(k)) &= \exp(z_{c})/\sum_{c \in \CMcal{C}_{S}} \exp(z_{c})\label{equ.PS}
    \\
    P^{(A| T_\text{set})}_{c,d(k)}(h(k)) &= \exp(z_{c,d(k)})/\sum_{c \in \CMcal{C}_{T_\text{set}}} \exp(z_{c,d(k)}),
\end{align}
with ${z}_c=u_{c,0} + u_{c,1}S(k)$, and $u_{c,0}$ and $u_{c,1}$ the parameters for the season softmax function.
For the time influence softmax function, we have $z_{c,d(k)}=\bfn{r}(h(k))\bfn{w}^T_{c,d(k)}$, and $\bfn{z}_{d(k)}=[z_{1,{d}(k)},\ldots,z_{|\CMcal{C}_{T_\text{set}}|,d(k)}]$, with parameters $\bfn{w}_{c,d(k)} \in \real^{2N}$.

Overall, the set of unknown parameters in the fraction of active households model (which we will learn in \cref{ss.learning}) is
\begin{equation}
\begin{split}
\Theta_{A}=\{&\eta_{s},u_{s,0},u_{s,1},\mu_{c,d},\bfn{w}_{c,d} \mid \\ & s \in \CMcal{C}_{S} , (c,d) \in \CMcal{C}_{T_\text{set}}\times \{1,2\}\}.
\end{split}
\end{equation}

%---------------------------------------------------------------------------
%              New Section
%---------------------------------------------------------------------------
% \vspace{-0.4cm}
\subsection{Thermodynamics for Single Building}
\label{sec:thermodynamics_single}

{
Similar to~\cite{Badings2020}, we derive a thermodynamics model for the temperature in buildings connected to the DHS.
In this section, \changes{we first derive a model for the temperature $T_b(k)$ in each building $b \in \mathbb{N}$,} which we then aggregate using the setpoint temperature $T_\text{set}$ and fraction of active households $A$.

We assume heat loss occurs through walls (${Q}_{\text{wall},b}$), windows (${Q}_{\text{win},b}$), and by ventilation (${Q}_{\text{vent},b}$).
Then, for a time discretization step $\tau \in \mathbb{R}$, we model the change $T_b(k+1) - T_b(k)$ in the indoor temperature of building $b$ within the time step $\tau$ as
\begin{equation}
\begin{split}
    T_b(k+1) = T_b(k) + \frac{\tau}{C_b} \big[
    &Q_{\text{space},b}(k) 
    - {Q}_{\text{wall},b}(k)
    \\
    &- {Q}_{\text{win},b}(k)
    - {Q}_{\text{vent},b}(k) \big],
    \label{equ.space_temperature_model}
\end{split}
\end{equation}
where $C_b$ is the thermal capacity of building $b$.
Following~\cite{Nielsen2006}, we now derive a model for the three loss terms in~\cref{equ.space_temperature_model}.
}

\paragraph*{Heat loss through walls}
Transfer ${Q}_{\text{wall},b}(k)$ through walls is proportional to the difference $T_b(k) - T_a(k)$ between indoor and ambient temperature at time $k$.
Denoting the solar radiation absorbed by the wall at time $k$ by $G_o(k)$, we obtain
\begin{equation}
{Q}_{\text{wall},b}(k) = K_{\text{wall},b} \big( T_{b}(k) - T_{a}(k) \big) -K_{\text{wall},b} \frac{\epsilon}{k_o}G_{o}(k),
\end{equation}%
where $K_{\text{wall},b} =(\frac{1}{k_w}+\frac{1}{k_i}+\frac{1}{k_o})^{-1}$, with $k_w$ the thermal conductivity of the walls, and with $k_i$ and $k_o$ the convection heat transfer coefficients of indoor and outdoor walls, respectively.
For simplicity, we assume that $k_w$, $k_i$, and $k_o$ are constant.
Moreover, $\epsilon$ is the fraction of $G_o$ not reflected by the wall, which is not a straightforward measurement; instead, we use the global radiance $G$ as an approximation of $G_o$.

\paragraph*{Heat loss trough windows}
Heat transfer ${Q}_{\text{win},b}(k)$ through windows occurs by conduction, convection, and solar radiation, resulting in the expression
\begin{equation}
{Q}_{\text{win},b}(k) = K_{\text{win},b} \big( T_{b} - T_{a}(k) \big) - \epsilon'_b G_{o}(k),
\end{equation}
where $K_{\text{win},b} =(\frac{1}{k_w'} + \frac{1}{k_i} + \frac{1}{k_o})^{-1}$, with $k'_{w}$ the window's thermal conductivity, and $k_i$ and $k_o$ are defined as for $Q_{\text{wall},b}(k)$.  

\paragraph*{Heat loss by ventilation}
Finally, the heat transfer by ventilation is again proportional to the difference $T_b(k) - T_a(k)$:
\begin{equation}
{Q}_{\text{vent},b}(k) =C\big( V_b + V_{w}(k) \big)\big( T_{b}(k) - T_{a}(k) \big),
\end{equation}
with $V_b$ the mechanical ventilation airflow (assumed to be constant), $V_w$ the airflow through the building, and $C$ the product of the specific heat capacity and mass density of air. 

{
\paragraph*{Condensed model}
Plugging in the definitions for ${Q}_{\text{wall},b}(k)$, ${Q}_{\text{win},b}(k)$, and ${Q}_{\text{vent},b}(k)$ in \cref{equ.space_temperature_model}, we obtain the condensed model (for a single building):
\begin{align}
    T_b(k+1) &= T_b(k) + \frac{\tau}{C_b} \Big[
        {Q}_{\text{space},b}(k) - \theta_{1,b} \big( T_b(k) - T_a(k) \big) 
        \nonumber
        \\
        &\enskip- \theta_{2,b} G(k) - \theta_{3,b} V_w(k) \big( T_b(k) - T_a(k) \big)
    \Big],
    \label{eq:Building_temp_condensed}
\end{align}
where we defined $\theta_{1,b} = K_{\text{wall},b} + K_{\text{win},b} + C V_b$,
$\theta_{2,b} = -K_{\text{wall},b} \frac{\epsilon}{k_o} - \epsilon_b'$, and
$\theta_{3,b} = C_b$ for brevity.
}

\subsection{Aggregated Space Heating Model}
\label{subsec:aggregated_space_heating}

Recall that the heat consumption at time $k$ is nonzero only for the fraction of active households $A(k)$, see \cref{equ.Ak}.
We assume that over a time horizon of $n \geq 1$ discrete time steps, the heat supply for the $B \in \mathbb{N}$ buildings in the DHS is perfectly balanced with the heat consumption, so we rewrite \cref{eq:Building_temp_condensed} as
\begin{align}
    \sum_{m=k}^{k+n}\! & \sum_{b=1}^B \widehat{Q}_{\text{space},b}(m) \! = \!\!\!\sum_{m=k}^{k+n}\! \sum_{j=1}^BA(m)\Big( 
     \theta_{1,j} \big( T_i(m) - T_a(m) \big) 
     \nonumber
    \\
    & + \theta_{2,j} G(m)  + \theta_{3,j} V_w(m) \big( T_i(m) - T_a(m) \big)
    \Big),
    \label{eq:Building_temp_balanced_group_B}
\end{align}
with $\widehat{Q}_{\text{space},b}(k)$ and $Q_{\text{space},b}(k)$ the approximate space heating supply and consumption of building $b$ at time $k$, respectively.

For the fraction $A(k)$ of buildings, we assume the value of $T_i(k)$ is relatively close to the temperature setpoint $T_\text{set}(k)$, such that we can approximate $T_i(k) = T_\text{set}(k)$. 
To account for the approximation error caused by assuming $T_i(k) = T_\text{set}(k)$ for all $k$, we introduce the backward shift operator $q^{-n}X(k) = X(k-n)$ and rewrite \cref{eq:Building_temp_balanced_group_B} as the following ARX model:
\begin{align}
    \overline{\CMcal{A}}_1(q^{-1})\widehat{Q}^{}_{\text{space}}(k) & =   A(k)\Big(\overline{\CMcal{B}}_1(q^{-1})\beta_1(T_\text{set}(k)-T_{a}(k)) 
    \nonumber
    \\ 
    & +  \overline{\CMcal{B}}_2(q^{-1})\beta_2G(k) 
    \\ 
    & + \overline{\CMcal{B}}_3(q^{-1}) \beta_3V_w(T_\text{set}(k)-T_{a}(k))\Big),
    \nonumber
    \label{eq:Building_temp_ARX}
\end{align}
where $\widehat{Q}^{}_{\text{space}}(k)=\sum_{b=1}^B \widehat{Q}_{\text{space},b}(k)$ is the approximated supply for $B$ buildings, $\beta_j=\sum_{b=1}^B \theta_{j,b}$ for $j=1,\ldots,3$, and $\overline{\CMcal{A}}_1(q^{-1})$ and $\overline{\CMcal{B}}_i(q^{-1})$ $i=1,\ldots,3$, are polynomials:
\begin{align}
\overline{\CMcal{A}}_1(q^{-1}) &= 1 - \overline{a}_{1,1}q^{-1} - \ldots - \overline{a}_{1,n_{a1}}q^{-n_{a1}} \\ 
\overline{\CMcal{B}}_i(q^{-1}) &= \overline{b}_{i,0} + \overline{b}_{i,1}q^{-1}, \ldots, \overline{b}_{i,n_{bi}} q^{-n_{bi}}.
    \label{eq:Building_temp_ARX_polys}
\end{align}
Expanding \cref{eq:Building_temp_ARX_polys}, we obtain the following functional form:
\begin{align}
\label{equ.full_space_heating}
\widehat{Q}_{\text{space}}&(k)\!  = \sum_{j=1}^{n_{a1}}\overline{a}_{1,j}\widehat{Q}_{\text{space}}(k-j) \\
& +A(k) \beta_1\!\!\sum_{j=0}^{n_{b1}}\overline{b}_{1,j} \big( T_\text{set}(k-j)-T_{a}(k-j) \big)   \nonumber\\
&+ A(k)\beta_2\!\!\sum_{j=0}^{n_{b2}}\overline{b}_{2,j} G(k-j)  \nonumber\\
&+ A(k)\beta_3\!\!\sum_{j=0}^{n_{b3}}\overline{b}_{3,j} V_w(k-j) \big( T_\text{set}(k-j)-T_{a}(k-j) \big). \nonumber
\end{align}
The polynomial orders $n_{a_1}$ and $m_{b_i}$ are hyper-parameters that can be set a priori or estimated from data.
Similar to~\cite{Nielsen2006}, we use dedicated filters for the ambient temperature, radiance and wind speed.
The parameters to be learned in \cref{equ.full_space_heating} are $\Theta_{\text{space}}=\{\overline{a}_{1,1},\ldots,\overline{a}_{1,n_{a1}},\overline{b}_{i,0},\ldots,\overline{b}_{i,n_{bi}},\beta_i \,\,|\,\, i \in \{1,2,3\}\}$.

%---------------------------------------------------------------------------
%              New Section
%---------------------------------------------------------------------------
{
\section{Modeling Hot Water Consumption in HELIOS}
}
\label{sec:Helios:hot_water}
We use the model proposed in \cite{task44_annex38} as the basis for modeling domestic hot water consumption. Let $Q_{\text{hot-water}}(k)$ be the hot water energy consumption at time $k$, is dependent on the users' consumption $U(k)$, along with a temperature correction factor:
\begin{align}
    Q_{\text{hot-water}}(k) &= U(k)\left[ 1 + \lambda \cos\left(\frac{2\pi (d_y(k)-d_{m}) }{365}\right)\right]\label{equ.hotwater_out}\\
    U(k)&=\sum_{b=1}^{B}\sum_{u \in \CMcal{C}_U} q_{b,u}(k) \label{equ.hotwater_U},
\end{align}
where $q_{b,u}(k)$ is the hot water energy demand for activity $u \in \CMcal{C}_U$ in building $b$, and $\CMcal{C}_U$ is the context set of activities requiring hot water. 
For example, $\CMcal{C}_{U}$ can contain periods of the day with different hot water demand, e.g., $\CMcal{C}_{T_\text{set}}=\{\text{sleeping},\text{waking-up},\text{working-hours},\text{after-work}\}$.
The parameter $\lambda$ represents the amplitude of the sine-curve for the yearly variation in the hot water energy demand. 
A typical value is $\lambda = 0.2$, indicating that hot water consumption is 20\% higher on colder (winter) days versus warmer (summer) days. 
Finally, $d_y(k)$ is the day of the year, and $d_{m}$ is the day of the year at which maximum hot water energy demand is expected. 

The contributions in \cref{equ.hotwater_U} are typically unknown, preventing us from estimating $Q_{\text{hot-water}}(k)$.
Instead, as shown by~\cref{fig.helios_causal}, we assume the user hot water demand depends only on the hour of the day and the type of day. Thus, we approximate $U(k)$ using a contextual mixture of experts model:
\begin{equation}
   \widehat{U}(k) = \sum_{u \in \CMcal{C}_U} P^{(U)}_{u,d(k)}(h(k))q_u,
   \label{equ.qhotwater}
\end{equation}
where constant $q_u$ is the maximum nominal hot water consumption for activity $u$ for all buildings $B$, and where $P^{(U)}_{u,d}(h(k))$ is the probability of activity $u$ being active at hour $h(k)$ and type of day $d$, which we define as the softmax function
\begin{equation}
P^{(U)}_{u,d(k)}(h(k)) = \exp(z_{u,d(k)})/\sum_{c \in \CMcal{C}_{T_\text{set}}} \exp(z_{u,d(k)}),\label{equ.PU}
\end{equation}
where $z_{u,d(k)}=\bfn{r}(h(k))\bfn{g}^T_{u,d(k)}$, and $\bfn{z}_{d(k)}=[z_{1,{d(k)}},\ldots,z_{|\CMcal{C}_{U}|,d(k)}]$, with $\bfn{g}_{u,d(k)} \in \real^{2N}$ the softmax vector parameters.

%---------------------------------------------------------------------------
%              New Paragraph
%---------------------------------------------------------------------------
% \subsubsection{Hot Water Model}
We combine the approximated user hot water consumption with the temperature correction factor to obtain the complete (approximated) hot water model:
\begin{align}
\widehat{Q}_{\text{hot-water}}(k) &= \widehat{U}(k)\left[ 1 + \lambda \cos\left(\frac{2\pi (d_y(k)-d_m)}{365}\right)\right]\label{equ.hotwater}.
\end{align}
The parameters to be estimated in the user hot water model are $\Theta_{U}=\{q_{u},\bfn{g}_{u,d} \mid u \in \CMcal{C}_{U} , (u,d) \in \CMcal{C}_{U}\times \{1,2\}\} $, and $\Theta_{\text{hot-water}}=\{\lambda\}$ in \cref{equ.hotwater}. 
%---------------------------------------------------------------------------
%              New Section
%---------------------------------------------------------------------------

\vspace{0.3cm}
\section{Modeling Piping Heat Loss in HELIOS}
\label{sec:Helios:heat_loss}
The supply, return, and ground temperatures influence the heat piping loss $Q_{\text{loss}}$.
Assuming that the supply and return pipes are laid side by side in the ground, the total heat loss is, according to~\cite{Frederiksen2013}, modeled as follows:
\begin{align}
    Q_{\text{loss}}(k)
            &= K_{e}\big( T_{sr}(k)-T_g(k) \big), %= \beta_4 \Delta T,
    \label{equ.heatloss_out}
\end{align}
where $K_e = 2(K_l-K_{sr})$, with $K_l$ and $K_{sr}$ the heat transfer coefficient between the supply and return pipes, and between the pipes and the ground, respectively.
Moreover, $T_{sr}=({T_s(k)+T_r(k)})/{2}$ is the equivalent piping temperature, where $T_s$, $T_r$, and $T_g$ are the supply, return, and ground temperatures, respectively. 
Measurements are available for the supply and return temperatures but not for the ground temperature, so we use the following model from~\cite{Banks2012,Calixto2021}:
\begin{align}
    T_{\text{g}}(k) = \overline{T_{a}} - &\Delta {T_{a}}\exp\left(  -z \sqrt{\frac{\pi}{365 \cdot \alpha_d}}\right) \times 
    \label{equ.groundemperature}
    \\ & \cos\left( \frac{2\pi}{365}\cdot (d(k)-d_{g}) -  z \sqrt{\frac{\pi}{365 \cdot \alpha_d}} \right),
    \nonumber
\end{align}
where $z$ is the pipe depth, $\overline{T_{a}}$ is the yearly average ambient temperature, $\Delta T_{a}$ is the amplitude difference between the annual maximum and minimum temperatures, $d_{g}$ is a shift in days related to the change in ground temperature due to the ambient temperature, and $\alpha_d$ is the ground thermal diffusivity. 
%$\alpha_d = 7 \times 10^{-7} \text{m}^2/\text{s}$ is the ground thermal diffusivity. 

%---------------------------------------------------------------------------
%              New Paragraph
%---------------------------------------------------------------------------
% \paragraph{Piping Heat Loss Model}
To model the heat loss, let $\widehat{Q}^{}_{\text{loss}}$ represent the heat supply lost in piping transmission, and ${Q}^{}_{\text{loss}}$ denote the piping heat loss. 
The relationship between the temperature change and the heat loss can be expressed as %\tb{I don't understand this; what is $\dot{T}_{sr}(k)$?}
\begin{equation}
  \dot{T}_{sr}(k) = \frac{1}{C_l} \left(\widehat{Q}^{}_{\text{loss}}(k) - {Q}^{}_{\text{loss}}(k) \right),
  \label{equ.loss_state}
\end{equation}
where $C_l$ is the thermal capacity of the piping system.
By discretizing equation \ref{equ.loss_state} at a time step \(\tau > 0\), we obtain:
\begin{equation}
  {T}_{sr}(k+1) = {T}_{sr}(k) + \frac{\tau}{C_l} \left(\widehat{Q}^{}_{\text{loss}}(k) - \beta_4 \big(T_{sr}(k)-T_g(k)\big) \right).
  \label{equ.loss_state_disc}
\end{equation}
Similar to the space heating model in Section \ref{sec:Helios:space_heating}, we assume a perfect balance between the heat supply and the total heat piping loss over a horizon of \(n\) time steps, which is written as
\begin{equation}
   \sum_{m=k}^{k+n} \widehat{Q}^{}_{\text{loss}}(m) = \sum_{m=k}^{k+n}\beta_4 (T_{sr}(m)-T_g(m)). 
   \label{equ.balance_loss}
\end{equation}
To account for the approximation error caused by assuming the heat loss to be equal to eq.\ \req{equ.heatloss_out}, we rewrite eq.\ \req{equ.balance_loss} as an ARX model of the following form:
\begin{equation}
\overline{\CMcal{A}}_2(q^{-1}) \widehat{Q}_{\text{loss}}(k) = \overline{\CMcal{B}}_4(q^{-1})\beta_4 \big(T_{sr}(k)-T_g(k)\big).
   \label{equ.heatloss_arx}
\end{equation}
\cref{equ.heatloss_arx} is rewritten in the extended form as
\begin{align}
\label{equ.heatloss_poly}
\widehat{Q}_{\text{loss}}(k+1) & = \sum_{j=1}^{n_{a2}}\overline{a}_{2,j}\widehat{Q}_{\text{loss}}(k-j)\nonumber 
\\
& + \sum_{j=0}^{n_{b4}} \overline{b}_{4,j}\beta_4 \big(T_{sr}(k-j)-T_g(k-j)\big).
\end{align}
The unknown parameters to be estimated in the loss model are $\Theta_{\text{loss}}=\{\overline{a}_{2,1},\ldots,\overline{a}_{2,n_{a2}},\overline{b}_{4,0},\ldots,\overline{b}_{4,n_{b4}},\beta_4 \}$. The hyper-parameters are the order of polynomials $n_{a2}, n_{b4}$.

%---------------------------------------------------------------------------
%              New Section
%---------------------------------------------------------------------------
\section{Learning the Parameters of the HELIOS Model}
% \subsection{The MAP algorithm}
\label{ss.learning}

% \subsection{The Complete HELIOS Model}
% \label{sec:Helios:complete}
The complete HELIOS model is obtained by combining the models for the space heating, the hot water, and the piping loss, which we have defined in the previous sections:
\begin{equation}
    Q({k}) = \widehat{Q}_{\text{space}}(k) + \widehat{Q}_{\text{hot-water}}(k) + \widehat{Q}_{\text{loss}}(k) + \epsilon(k),
    \label{equ.finalmodel}
\end{equation}
where $\epsilon$ is a white Gaussian noise that captures unmodeled disturbances.
To fit the model in \cref{equ.finalmodel}, we need to estimate a number of parameters, namely $\Theta =\{\Theta_{T_\text{set}},\Theta_{A},\Theta_{U},\Theta_{\text{space}},\Theta_{\text{hot-water}},\Theta_{\text{loss}}\}$, where the respective parameter are defined in~\cref{sec:Helios:space_heating,sec:Helios:hot_water,sec:Helios:heat_loss}.
Moreover, we need to define the contextual information (expert domain knowledge) to integrate the contextual models defined in \cref{equ.Ts,equ.alpha_s,equ.alpha_i,equ.qhotwater} into the HELIOS model.

To estimate the unknown parameters of the HELIOS model, we use \emph{maximum a posteriori estimation} (MAP) to integrate prior knowledge of parameters efficiently. 
In this section, we often omit the sampling index $k \in \mathbb{N}$ for brevity.

%---------------------------------------------------------------------------
%              New Subsection
%---------------------------------------------------------------------------
\subsection{HELIOS distribution}
The complete conditional posterior $p_H$ of the HELIOS model is expressed by the following decomposition of the causal graph heat load model shown in Fig.\ \ref{fig.helios_causal}:
\begin{equation}
    p_H =  p_{Q}p_{Q_{\text{s}}}p_{Q_{\text{w}}}p_{Q_{\text{l}}},
    \label{equ.pH}
\end{equation}
where each factor is defined as
\begin{align}
&p_{Q} = p(Q \mid Q^{}_{\text{space}}, Q^{}_{\text{hot-water}}, Q^{}_{\text{loss}}) \label{eq.pq} \\
&p_{Q_{\text{s}}}\! = \!p(Q^{}_{\text{space}} \!\! \mid \!\! T^{}_{\text{set}},A^{},T_a,V_w,G)  p(T^{}_{\text{set}} \!\! \mid \!\!h,d)p(A^{} \!\! \mid \!\! S,h,d)\label{eq.pqspace} \\
&p_{Q_{\text{w}}} = p(Q^{}_{\text{hot-water}} \mid h,d) p(U \mid h,d) \label{eq.pqhotwater} \\
&p_{Q_{\text{l}}} = p(Q_{\text{loss}} \mid T_s,T_r,T_a)\label{eq.pqloss}.
\end{align}
The unmeasured variables $Q^{}_\text{space}, Q^{}_\text{hot-water}, Q^{}_\text{loss}, A^{}$, $T^{}_{\text{set}}$ and $U$ are unknown, so we cannot compute the likelihood of distribution \cref{equ.pH} directly. 
Instead, we work with the incomplete distribution by marginalizing \cref{equ.pH} over $A^{}$, $T^{}_{\text{set}}$ and $U$, and by approximating $Q^{}_m$, for $m \in \{\text{space}, \text{hot-water}, \text{loss}\}$ from the heat balance of the system.

From \cref{equ.finalmodel}, we compute the residual obtained from predicting the heat load $Q$ while only considering the contribution of the estimated mean of $Q^{}_m$ for all $m$. For example, the true space heating is approximated as 
% \begin{equation}
$
{Q}^{}_{\text{space}} = Q - \widehat{Q}_{\text{hot-water}} - \widehat{Q}_{\text{loss}}.
$
% \end{equation}
%
As we cannot estimate the quantities $A^{}$, $T^{}_{\text{set}}$ and $U^{}$ as we did for the $Q^{}_m$'s, we marginalize \cref{eq.pq} over the unmeasured variables, corresponding to the incomplete data distribution. Therefore, the incomplete distribution is given by
\begin{align}
    \overline{p}_{H}  &=  \int_{A^{}} \int_{T^{}_{\text{set}}}\!\int_{U^{}}  p_{Q}{p}_{Q_{\text{s}}}{p}_{Q_{\text{w}}}p_{Q_{\text{l}}}dA^{}dT^{}_{\text{set}}dU^{}\nonumber\\
    &= p_{Q}p_{Q_{\text{l}}}\int_{A^{}} \int_{T^{}_{\text{set}}}{p}_{Q_{\text{s}}}dA^{}dT^{}_{\text{set}}\int_{U^{}}  {p}_{Q_{\text{w}}}dU^{}.
    \label{equ.ptheta}
\end{align}
Computing the marginalized distribution $\overline{p}_{Q_{\text{s}}}=\int_{A^{}}\int_{T^{}_{\text{set}}}{p}_{Q_{\text{s}}}dA^{}dT^{}_{\text{set}}$ involves integrating over variables $A$ and $T_{\text{set}}$.
Thus, we define the conditional distributions of the unmeasured variables $A$ and $T_{\text{set}}$. 
The conditional distributions of $T_{\text{set}}$ given $h$, and $d$, and $A$ given $S$, $h$, $d$, are expressed as:
\begin{align}
&p(T^{}_{\text{set}} \mid h,d,\Theta_{\text{set}}) = \sum_{c \in \CMcal{C}_{T_\text{set}}} P^{(T_\text{set})}_{d,c}(h)p(T^{}_{\text{set}} \mid\zeta_{d,c})\label{equ.probTset}\\
&p(A^{} \mid S, h,d,\Theta_{\text{A}}) =\sum_{c \in \CMcal{C}_S}^{} P^{(A| S)}_c(S) p(A^{} \mid \eta_{c})\label{equ.probAk}\\
&\hspace{80pt}\times\sum_{c\in \CMcal{C}_{T_\text{set}}}^{} P^{(A| T_\text{set})}_{c,d}(h) p(A^{} \mid\mu_{c,d}).\nonumber
\end{align}
%Here, \cref{equ.probTset,equ.probAk} represent the conditional probabilities of $T_{\text{set}}$ and $A$, respectively. 
The marginalized distribution $\overline{p}_{Q_{\text{s}}}$ is then computed as follows:
\begin{align}
\overline{p}_{Q_{\text{s}}} = & \sum_{i \in \mathcal{C}_S} \sum_{j \in \mathcal{C}_{T_{\text{set}}}} \sum_{l \in \mathcal{C}_{T_{\text{set}}}}P^{(A| S)}_i(S)   P^{(A| T_\text{set})}_{j,d}(h)  P^{(T_\text{set})}_{l,d}(h) \nonumber \\
&\times p(Q^{}_{\text{space}}| \widehat{Q}^{}_{\text{space}\mid A^{}=\eta_c\mu_{c,d}, T^{}_{\text{set}}=\zeta_{d,c}}, T_a, V_w, G, \Theta_{\text{space}} ) \nonumber\\
&\times p({A^{}}=\eta_c\mu_{c,d} \mid S, h,d) p(T^{}_{\text{set}}=\zeta_{d,c}\mid h,d),\label{equ.margpQs}
\end{align}
\cref{equ.margpQs} involves summing over different contexts, namely $\mathcal{C}_S$, and $\mathcal{C}_{T_{\text{set}}}$. 
Variable $Q_{\text{space}}$ is conditioned on specific values of the parameters $T_a$, $V_w$, and $G$.
Also, $\widehat{Q}^{}_{\text{space}}$ is fixed to the values of $A^{}=\eta_c\mu_{c,d}$ and $T^{}_{\text{set}}=\zeta_{d,c}$. The terms $p(A=\eta_c\mu_{c,d} \mid S, h, d)$ and $p(T_{\text{set}}=\zeta_{d,c} \mid h, d)$ represent the priors of the parameters $\eta_c$, $\mu_{c,d}$, and $\zeta_{d,c}$. Since these priors do not depend on $S$, $h$, and $d$, they can be rewritten as $p(\eta_c)$, $p(\mu_{c,d})$ and $p(\zeta_{d,c})$, respectively, or in the equivalent form $p(\Theta_{T_{\text{set}}})$, and $p(\Theta_{A})$.

Similarly, to compute the marginalizing distribution of \cref{eq.pqhotwater} over $U^{}$, defined as $\overline{p}_{Q_{\text{w}}} = \int_{U^{}} {p}_{Q_{\text{w}}}  dU^{}$, we first define the distributions of the unmeasured variable $U^{}$ as
\begin{align}
&p(U \mid h,d,\Theta_U) = \sum_{u \in \CMcal{C}_U} P^{(U)}_{u,d}(h)p(U^{}\mid q_u).
\label{eq.pUstar}
\end{align}
Marginalizing \cref{eq.pqhotwater} over $U^{}$, as $\overline{p}_{Q_{\text{w}}} = \int_{U^{}} {p}_{Q_{\text{w}}}  dU^{}$, we obtain
\begin{align}
\overline{p}_{Q_{\text{w}}} = & \sum_{u \in \CMcal{C}_U} P^{(U)}_{u,d}(h)\label{equ.margpQw} \\
& \times P (Q^{}_{\text{hot-water}} \mid \widehat{Q}^{}_{\text{hot-water} \mid U=q_u}, d) p(U=q_u \mid h,d). \nonumber
\end{align}
In \cref{equ.margpQw}, we compute the marginalized distribution $\overline{p}_{Q{\text{w}}}$ of the hot water $Q_{\text{hot-water}}$ by summing over the contextual information sets $u \in \mathcal{C}_U$. The term $p(U=q_u \mid h, d)$ represents the prior of parameter $q_u$, and can be rewritten as $p(q_u)$ as it is independent of $h,d$, or in the equivalent form $p({\Theta_{\text{hot-water}}})$.
Hence, the distributions in \cref{eq.pq,eq.pqloss} are defined as
\begin{align}
p_Q &= p(Q|\widehat{Q}^{}_{\text{space}}+\widehat{Q}^{}_{\text{hot-water}}+\widehat{Q}^{}_{\text{loss}})\\
p_{Q_l} &= p(Q^{}_{\text{loss}}|\widehat{Q}^{}_{\text{loss}}, T_s, T_g,\Theta^{}_{\text{loss}})
\end{align}
The priors of all parameters can be summarized as the set $\Theta = \{\Theta_{T_{\text{set}}},\Theta_{A},\Theta_{\text{space}},\Theta_{\text{hot-water}},\Theta_{\text{loss}} \}$.

\subsection{Learning \& Knowledge Integration}
\label{ss.knowledge}
Classically, we would maximize the posterior of the incomplete distribution over a set of examples $k$, i.e., $\CMcal{L}(\Theta) = \prod_k \overline{p}_H(k)p(\Theta)$.
However, in this paper, 
we intend to integrate expert knowledge expressed by the contextual models~\cite{Souza2022}, which we used in \cref{equ.Ts,equ.alpha_s,equ.alpha_i,equ.qhotwater}. While priors provide prior knowledge of model parameters,  contextual models further enhance the model's expected behavior in known situations. 
The idea is to maximize the weighted likelihood and optimize them accordingly (for details on this approach, we refer to \cite{Souza2022}). 
Specifically, we maximize the weighted posterior of the incomplete distribution:
\begin{align}
\CMcal{L}_w(\Theta) = \prod_k \overline{p}^w_{H}(k)p(\Theta),
\label{equ.weightedMAP}
\end{align}
where $\overline{p}^w_{H} = p_{Q}p_{Q_{\text{l}}} \overline{p}^w_{Q_{\text{s}}}\overline{p}^w_{Q_{\text{w}}}$, with $\overline{p}^w_{Q_{\text{s}}}$ as the weighted likelihood of the space contribution, and $\overline{p}^w_{Q_{\text{w}}}$ as the weighted likelihood of hot water contribution. Let $\pi^{(A| T_\text{set})}_{i}, \pi^{(A| S)}_{j}$ be the weights of context $i \in \CMcal{C}_S$ and $j \in \CMcal{C}_{T_\text{set}}$ for the fraction of active households given the season and the temperature setpoint, respectively, $\pi^{(T_\text{set})}_{l}$ be the weights for the temperature set-points context $l \in \CMcal{C}_{T_\text{set}}$, and $ \pi^{(U)}_{u}$ be the weights of the hot water consumption context $u \in \CMcal{C}_U$. Therefore, we define $\overline{p}^w_{Q_{\text{s}}}$
\begin{align}
\overline{p}^w_{Q_{\text{s}}} = & 
\!\!\!\!\!\!\!\!
\sum_{ \substack{ i \in \CMcal{C}_S, j\in \CMcal{C}_{T_\text{set}}, l\in \CMcal{C}_{T_\text{set}}}}
\!\!\!\!\!\!\!\!\!\!\!\!\!
\Big[\!P^{(A| S)}_i(S)\!\Big]^{\!\textsuperscript{$\pi^{(A| T_\text{set})}_{i}$}} 
\!\!\!\!\!\!\!\!\!  
\Big[\!P^{(A| T_\text{set})}_{j,d}(h)\!\Big]^{\!\textsuperscript{$\pi^{(A| S)}_{j}$}}
\!\!\!\!\!\!\!\!\!   
\Big[\!P^{(T_\text{set})}_{d,l}(h)\!\Big]^{\!\textsuperscript{$\pi^{(T_\text{set})}_{l}$}}\nonumber \\
&\times p(Q^{}_{\text{space}}|\widehat{Q}^{}_{\text{space} \mid A=\eta_i\mu_{j,d}, T_{\text{set}}=\zeta_{l,d}}, T_a, V_w,G)^{\pi_{ijl}} \nonumber\\
&\times p(\eta_i)p(\mu_{j,d}) p(\zeta_{l,d}).
\end{align}
where $\pi_{ijl}=\pi^{(A| T_\text{set})}_{i}\pi^{(A| S)}_{j} \pi^{(T_\text{set})}_{l}$. Similarly, for the hot water distribution, we have
\begin{align}
\overline{p}^w_{Q_{\text{w}}} =& \sum_{u \in \CMcal{C}_U} \Big[P^{(U)}_{u,d}(h) \Big]^{\pi^{(U)}_{u}} \\
&\times P (Q^{}_{\text{hot-water}} \mid \widehat{Q}^{}_{\text{hot-water} \mid U=q_u} ,d)^{\pi^{(U)}_{u}} p(q_u ).
\end{align}

In the learning process, we iterative update the parameters $\Theta$, maximize the posterior of the incomplete distribution \cref{equ.weightedMAP}, and re-update unmeasured variables and re-estimate the new parameters. This leads to the following iterative scheme:
\begin{enumerate}
\item Initialize $\Theta$;
\item Compute $Q_m$ with the values of $\Theta_{}$ at the current iteration, find the MAP estimation as
% \begin{equation}
   $\Theta = \argmax_{\Theta^*} \CMcal{L}_w(\Theta^*)$, and
   % \label{equ.map_theta}
% \end{equation}
Update $Q_m$ for all $m \in \{\text{space},\text{hot-water},\text{loss}\}$ with the estimated values of $\Theta$;
\item Repeat this procedure until all parameters in $\Theta$ converge.
\end{enumerate}

%---------------------------------------------------------------------------
%              New Section
%---------------------------------------------------------------------------
%\tb{Should the colors in Fig 4a be reversed? I think so..}

\section{Experimental Results}
\begin{figure}[!tb]
\centering
\includegraphics[width=.9\columnwidth]{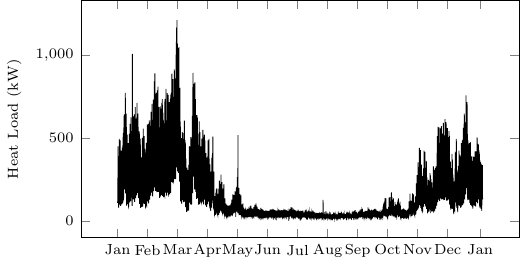}\caption{Hourly heat demand (in kW) for the year of 2018.}
\label{fig.fig_heat}
\vspace{-0.3cm}
\end{figure}

We test the HELIOS model in a case study to forecast heat demand in a DHS substation located in the Netherlands using hourly data shown in \cref{fig.fig_heat}. The weather variables, ambient temperature, solar radiation, and wind speed are obtained from the Dutch meteorological institute KNMI.
\relax

\paragraph{Experimental setup} We compare HELIOS with various \changes{statistical models widely employed in modeling the heat load in DHS. We choose }. 
First, we consider several static representation models, namely linear regression (LR), LR with LASSO penalty (LASSO), LR with elastic net penalty (EN), linear regression with ridge regression penalty (RR), support vector regression (SVR), and XGboost. 
Second, we consider several dynamic prediction models, namely ARMA, ARIMA, ARMAX, ARIMAX, LSTM, and RNN. \changes{We include both dynamic and static models to cover the spectrum present in the literature, allowing us to benchmark the HELIOS against commonly reported models.}
{
The input variables for all models include the outdoor (ambient) temperature $T_a$, wind velocity $V_w$, solar radiation $G$, the type of day $d$, and the Fourier variable \cref{equ.hourToVec} for the hour of the day $h$.}
% \begin{table}[!t]
% \caption{Input variables for HELIOS (and all alternative models).}
% \label{tab.inputs}
% {\scriptsize
% \begin{center}
% \begin{tabular}{cl}
% \toprule
%      Variable & Description \\ 
% \hline 
%      ${T}_a$    &  Air temperature \\ 
%      ${V_w}$    &  Wind speed \\ 
%      ${G}$      &  Solar radiation \\ 
%      $h$        &  Hour of the day \\
%      $d$        & Type of day (weekend, weekday) \\
% \bottomrule
% \end{tabular}
% \vspace{-1.5em}
% \end{center} }
% \end{table}\relax
All models were trained using data for the year 2018 and tested using data for 2019. 
The hyper-parameters are selected based on the cross-validation error in the training data. \changes{
For RNNs and LSTMs, we tuned the number of units within the range $[32, 512]$ and the number of layers between $[1, 5]$ using hold-out cross-validation. For ARX, ARMAX, and ARIMAX models, we adjusted lags within $[1, 48]$, moving average components within $[1, 12]$, and difference orders within $[1, 2]$, with parameters selected using the Bayesian Information Criterion (BIC). For LASSO, Elastic Net, and Ridge Regression, we tuned regularization parameters from the range $[0.001, 10]$ and employed 10-fold cross-validation. For SVR, we adjusted the regularization parameter within $[0.1, 100]$ and tube size within $[0.001, 1]$ using 10-fold cross-validation. For XGBoost, we conducted a search over maximum depth within $[2, 6]$ and estimators within $[50, 200]$ using 10-fold cross-validation.}
We compare forecasting performance by the root mean square error (RMSE), the coefficient of determination ($\text{R}^2$), the mean absolute error (MAE), and the mean absolute percentage error (MAPE). 

%---------------------------------------------------------------------------
%              New Section
%---------------------------------------------------------------------------

\subsection{HELIOS Parameters }
\cref{tab.parameters_helios} presents a full overview of the inputs required for the HELIOS model.
We discuss the values of the hyper-parameters, contexts, and priors we use in our experiments.
We write a normal distribution with mean $\mu$ and variance $\sigma$ as $\text{normal}(\mu,\sigma)$, and a half-normal distribution as $\text{half-normal}(\mu,\sigma)$.

{\subsubsection{Hyper-parameters in HELIOS} 
In the experiment, we use $P=3$ as the number of Fourier variables.
%$d = \{1,2\}$ to distinguish between weekdays ($d=1$) and weekends/holidays ($d=2$).
For the space heating model (\cref{equ.full_space_heating}) and heat loss model (\cref{equ.heatloss_poly}), we define the polynomial orders as $n_{a1}=1, n_{bi}=1, n_{a2}=1, n_{b4}=1$, thus using a first-order model to reduce computational demands.
Using higher orders may improve the model accuracy but also lead to increased complexity and a higher risk of overfitting.

}

\begin{table}[!t]
\caption{Hyper-parameters, contexts, and priors required for HELIOS.}
\vspace{-0.2cm}
\label{tab.parameters_helios}
{\scriptsize
\begin{center}
\begin{tabular}{clc}
\toprule
 Hyper-parameters & Description \\ 
\hline
$P = 3$ & Number of Fourier components\\
\multirow{2}{*}{$n_{a1}=1, n_{bi}=1$}  &  Order of polynomials of \\&space heating model $i \in \{1,2,3\}$;   \\ 
$n_{a2}=1, n_{b4}=1$  &  Order of polynomials of piping loss model;   \\ 
\toprule
     Contexts & Description \\ 
\hline 
 $\CMcal{C}_{T_\text{set}}$  &  Context for temperature setpoint model;   \\ 
 $\CMcal{C}_{S}$  &  Context for season model;   \\ 
 $\CMcal{C}_{U}$  &  Context for user behavior model;  \\ 
 \toprule
     Priors & Description \\ 
\hline 
$p(\bfn{z}_{c,d})$ & Prior for gates parameters of temperature setpoint model;\\ 
\multirow{2}{*}{$p(\bfn{w}_{c,d})$}  & Prior for gates parameters of temperature setpoint\\& influence  on the fraction of active households model;\\ 
$p(\bfn{g}_{c,d})$  & Prior for gates parameters of user consumption model;\\ 
\multirow{2}{*}{$p(u_{s,0}), p(u_{s,1})$}  & Prior for gates parameters of season influence \\& on the fraction of active households model;\\ 
\multirow{2}{*}{$p(\zeta_{c,d})$}  & Prior for the temperature setpoint of context $c \in \CMcal{C}_{T_\text{set}}$ \\&at type of day $d$.\\ 
$p(\eta_{c}) $  & Prior for the effect of season on the fraction of active\\& households for $c \in \CMcal{C}_{S}$.\\ 
$p(\mu_{c,d}) $  & Prior for the effect of temperature setpoint on the fraction\\& of active households for $c \in \CMcal{C}_{T_\text{set}}$ at the type of day $d$\\ 
\multirow{3}{*}{$p(\beta_{i})$}  & Priors for temperature, solar radiance, wind speed, \\&and supply/return temperature influence on heat load \\&$i\in \{1,2,3,4\}$ \\
\multirow{4}{*}{$p(\overline{a}_{l,m}), p(\overline{b}_{i,j})$}  & Priors for temperature, solar radiance, wind speed, \\&and supply/return temperature influence on heat load \\&$l\in \{1,2\}$, $m\in \{1,\ldots,n_{al}\}$, \\&$i\in \{1,2,3,4\}$, $j\in \{1,\ldots,n_{bi}\}$ \\
\multirow{2}{*}{$p(\lambda)$}   & Prior of the parameter that indicates the amplitude \\&gain of hot water in winter, related to summer\\
\multirow{2}{*}{$p(q_u)$}   & Prior for maximum nominal hot water consumption at\\& activity $u \in \CMcal{C}_U$\\
\bottomrule
\end{tabular}
\vspace{-1.5em}
\end{center} }
\end{table}\relax

%---------------------------------------------------------------------------
%              New Subsubsection
%---------------------------------------------------------------------------
\subsubsection{Contexts in HELIOS}
\begin{figure}[!t]
\centering
\subfigure[ ]
{\label{fig.fig_context_indoor_temperature}\includegraphics[width=.3\columnwidth]{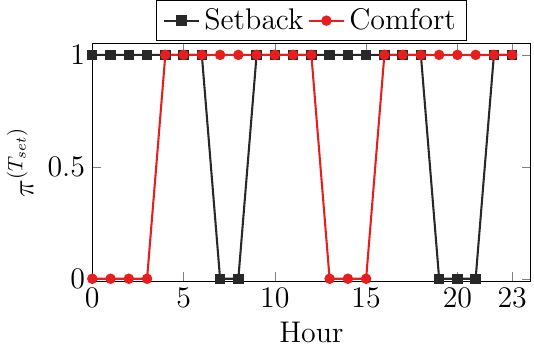}  }
\subfigure[ ]
{\label{fig.fig_context_season}\includegraphics[width=.3\columnwidth]{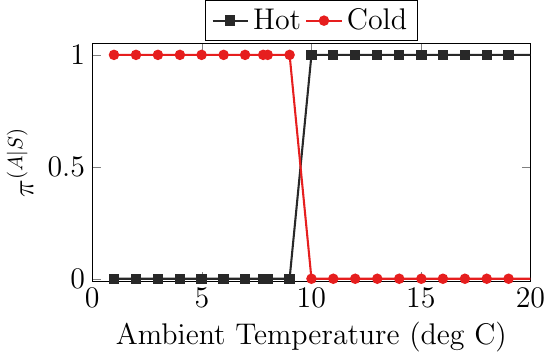}  }
\subfigure[ ]
{\label{fig.fig_context_user}\includegraphics[width=.3\columnwidth]{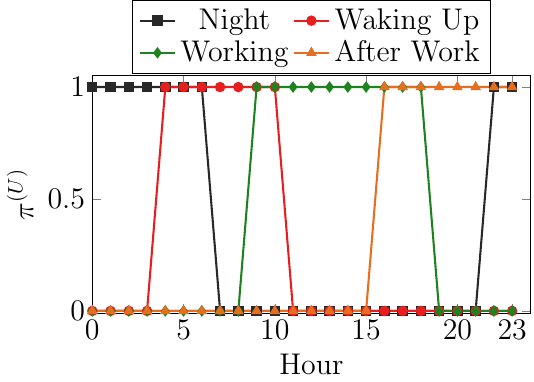}  }
\subfigure[ ]
{\label{fig.fig_gate_indoor_temperature}\includegraphics[width=.3\columnwidth]{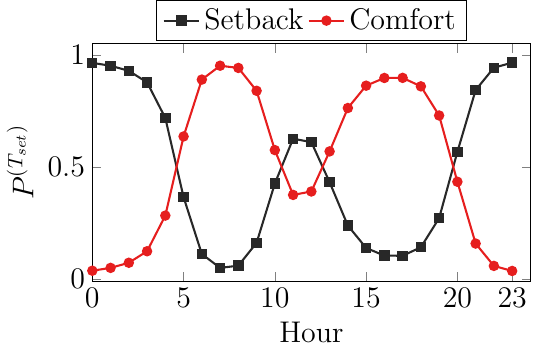}  }
\subfigure[ ]
{\label{fig.fig_gate_season}\includegraphics[width=.3\columnwidth]{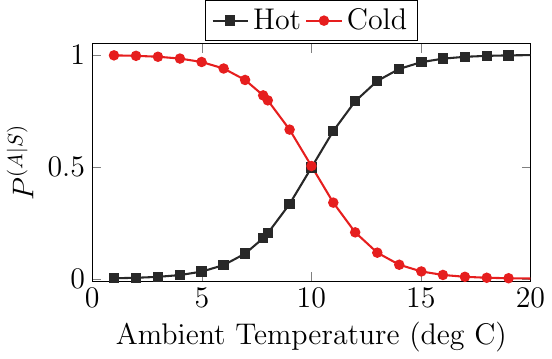}  }
\subfigure[ ]
{\label{fig.fig_gate_user}\includegraphics[width=.3\columnwidth]{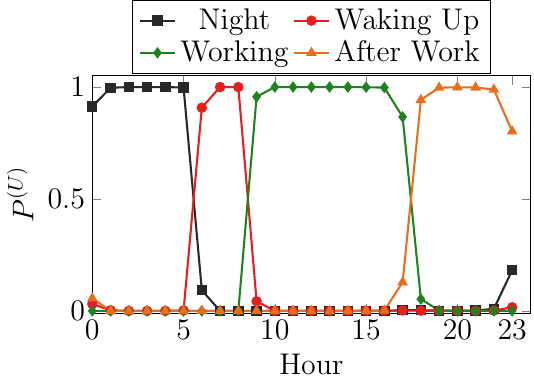}  }
\caption{Contexts for a) the temperature set-point, b) the user hot water consumption, and c) the season influence, and the respective fitted probabilities for d) the temperature set-point, e) the user hot water consumption, and f) the season influence } 
\label{fig.contexts_train}
\vspace{-0.4cm}
\end{figure}

The contextual information used by HELIOS in this experiment, together with the probabilities \cref{equ.PTs,equ.PS,equ.PU}, are shown in \cref{fig.contexts_train}. 
We discuss the contexts for each of these three elements in more detail.

\paragraph*{Temperature setpoint}
We define two contexts for the temperature setpoint: \emph{setback} (between 10 pm--6 am and 9 am--6 pm) and \emph{comfort} (between 4 am--10 am and 4 pm--12 am). 
Importantly, the overlap between these contexts indicates our uncertainty or limited knowledge about the precise context at specific times, as shown in \cref{fig.fig_context_indoor_temperature}. The fitted contexts, represented by the probability for the temperature setpoint (\cref{equ.PTs}), are displayed in \cref{fig.fig_gate_indoor_temperature} and demonstrate a strong correspondence to the expected context behavior.

\paragraph*{Season}
As shown in \cref{fig.fig_context_season}, we define two seasonal contexts: \emph{hot} (temperatures above 10\textdegree C) and \emph{cold} (below 10\textdegree C). %The hot context included temperatures above 10\textdegree C, while the cold context represented temperatures below 10\textdegree C. 
The corresponding fitted probabilities, calculated using \cref{equ.PS}, are shown in \cref{fig.fig_gate_season}. 
Notably, the fitted probabilities exhibit distinct seasonal profiles for both the hot and cold contexts.

\paragraph*{Hot water consumption}
As shown in \cref{fig.fig_context_user}, we define four contexts for the hot water demand: \emph{night} (between 10 pm--6 am), \emph{waking up} (4 am--10 am), \emph{working hours} (9 am--6 pm), and \emph{after work} (4 pm--12 am). 
The fitted context, represented by the probability function in \cref{equ.PU}, accurately captures the user contexts within these periods, as observed in \cref{fig.fig_gate_user}.
\relax
The results indicate that the contexts were successfully integrated into the HELIOS model. 
Analyzing the results of fitted contexts is critical to understanding the model. 
If there is a significant discrepancy between the provided context and the model output, the model may need to be retrained or the underlying cause of the divergence identified.

%---------------------------------------------------------------------------
%              New Subsubsection
%---------------------------------------------------------------------------
{
\subsubsection{Priors in HELIOS}
Integrating priors within HELIOS helps to guide parameter selection during the MAP estimation.

\paragraph*{Gates} We adopt a weakly informative prior stance for the gate parameters.
Specifically, we use zero-mean Gaussian priors $p(\bfn{z}_{c,d}), p(\bfn{w}_{c,d}),p(\bfn{g}_{c,d}), p(\bfn{w}_{c,d}), p(u_{s,0}), p(u_{s,1}) \sim \text{normal}(0, 2)$ for the gate parameters.

\paragraph*{Temperature setpoint} 
Recall that we consider two contexts for the temperature setpoint $\zeta_{c,d}$, such that $\CMcal{C}_{T_\text{set}} = \{\text{setback}, \text{comfort}\}$.
For the setback context $c=2$, we assume the Gaussian priors $p(\zeta_{\text{setback},1}), p(\zeta_{\text{setback},2})\sim \text{normal}(16^{\circ}\text{C},2^{\circ}\text{C})$.
For the comfort context (temperature), we assume priors$p(\zeta_{\text{comfort},1}), p(\zeta_{\text{comfort},2}) \sim \text{normal}(20^{\circ}\text{C},2^{\circ}\text{C})$.
These prior correspond with the findings presented in \cite{Santin2017}.

\paragraph*{Fraction of active households}
For the effect of the season contexts $\mathcal{C}_S = \{\text{hot}, \text{cold}\}$ on the fraction of active households, we use half-normal priors $\eta_{\text{hot}} \sim \text{half-normal}(0, 0.1)$ and $\eta_{\text{cold}} \sim \text{half-normal}(1, 0.1)$.
Here, $\eta_{\text{hot}}$ and $\eta_{\text{cold}}$ are restricted to the interval $[0, 1]$. 
For the unit simplexes $\mu_{c,d(k)}$, with $c \in \mathcal{C}_{T_{\text{set}}}$ and type of day $d$ (used in \cref{equ.alpha_i}), we use $p(\mu_{\text{setback},1}), p(\mu_{\text{setback},2}) \sim \text{half-normal}(0.2, 0.2)$ and $p(\mu_{\text{comfort},1}), p(\mu_{\text{comfort},2}) \sim \text{half-normal}(0.8, 0.2)$, with all parameters constrained to the interval $[0, 1]$.

\paragraph*{Space heating and piping loss}
The space heating model parameters capture the influence of ambient temperature, wind speed, and supply/return temperature on the heat load and consist of the static parameter $p(\beta_i) \sim \text{half-normal}(0, 10)$ and dynamic parameters $p(\overline{a}_{l,m}), p(\overline{b}_{i,j}) \sim \text{half-normal}(0, 1)$, both considered weakly informative priors.

\paragraph*{Hot water consumption}
For the hot water consumption model, weakly informative priors are assumed for the maximum nominal hot water use for activity $u \in \mathcal{C}_U$, with $p(q_u) \sim \text{half-normal}(0, 10)$. 
Specifically, we choose the prior $p(\lambda) \sim \text{half-normal}(0.2, 0.05)$ to approximate a $20\%$ increase in hot water consumption during winter compared to summer.

}

\begin{figure}[!t]
\includegraphics[width=1\columnwidth]{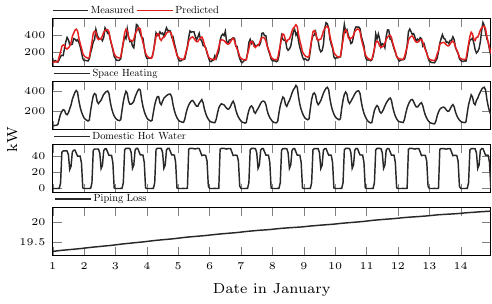}
\vspace{-0.8cm}
\caption{Demand predicted by HELIOS vs. measured demand for Jan.\ 2019.}
\label{fig.fig_forecasting_january2019}
\end{figure}\relax
\begin{figure}[!t]
\includegraphics[width=1\columnwidth]{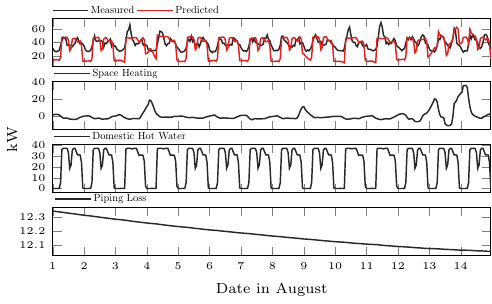}
\vspace{-0.8cm}
\caption{Demand predicted by HELIOS vs. measured demand for Aug.\ 2019.}
\label{fig.fig_forecasting_august2019}
\end{figure}\relax

\vspace{-0.3cm}
%---------------------------------------------------------------------------
%              New Subsection
%---------------------------------------------------------------------------
{
\subsection{The importance of identifying contexts}
\label{ss.evaluating_helios}

\changes{
To evaluate the importance of the contextual information, we compare three scenarios for the contexts $\CMcal{C}_{T_{\text{set}}}$,  $\CMcal{C}_{S}$, $\CMcal{C}_{U}$: (i)~HELIOS-NC, with no contextual information at all (i.e., all context distributions equal to one); (ii)~HELIOS-WC, with randomly assigned contexts (respecting the constraint defined by the possibility distribution, see \cref{sec:Helios:ExpertModels}); and (iii)~HELIOS, where the contexts are defined by the experts. The performance results are presented in \cref{tab.helios_evaluation}.} 

\changes{The best performance (in all metrics) is obtained for HELIOS with the correct contexts. 
HELIOS-WC performs worse in all metrics, highlighting the negative impact of assigning random contexts. On the other hand, HELIOS-NC performs better than HELIOS-WC but still falls short of the accuracy achieved with the correct contexts, as evidenced in the performance metrics, for example, the MAPE drops from 28.49\% in HELIOS-NC and 38.01\% in HELIOS-WC to 20.08\% compared to HELIOS.
Thus, providing incorrect contexts reduces the prediction quality of the HELIOS model.}

\changes{Besides performance, the transparency offered by the HELIOS model allows for a detailed inspection of the context's impact on the results. %enabling the acceptance of the model if it behaves as expected. 
This, in turn, enhances the trustworthiness of the model outcomes. 
In \cref{ss.discussion}, we explore how the disaggregated components facilitate the visual validation of the expected heat load profiles for the summer and winter seasons.}

\begin{table}[!t]
\scriptsize
\caption{HELIOS performance with different context settings for predicting the head demand over a horizon of 12 hours.}
\centering
\begin{tabular}{lrrr}
\toprule
{} &     HELIOS-NC& HELIOS-WC & HELIOS \\
\midrule
$\text{R}^2$  & 0.86 &     0.84 &  0.87 \\
RMSE  & 55.97 &    59.85 &  52.36 \\
MAE  & 35.81  &    40.77 &  25.88  \\
MAPE  & 28.49 &    38.01 &  20.08  \\
\bottomrule
\end{tabular}
\label{tab.helios_evaluation}
\vspace{-0.4cm}
\end{table} 

}

\vspace{-0.3cm}
\subsection{Comparing HELIOS with existing prediction models}
\cref{fig.fig_forecasting_january2019,fig.fig_forecasting_august2019} present the heat demand for January 2019 {and August 2019, in winter and summer seasons, respectively}. The red line shows the demand predicted by HELIOS, which accurately matches the actual demand (the black line). In January (\cref{fig.fig_forecasting_january2019}), over 90\% of the heat load is allocated to space heating, as anticipated for a cold month. {In August, the heat load is mostly domestic hot water, while the space heating is zero for most of the period, as expected. On the 13th and 14th of August, the temperature dropped to around 11$\,^\circ $C, gating energy use for space heating, as observed in \cref{fig.fig_forecasting_august2019}.}

\begin{table*}[!t]
\scriptsize
\setlength{\tabcolsep}{3.6 pt}
\caption{Performance metrics for predicting the head demand over a horizon of 12 hours in the test set.}
\centering
\begin{tabular}{lrrrrrrrrrrrrrrr}
\toprule
{} &     LR & XGBoost & RR & EN & LASSO & SVR &   ARMA &  ARIMA &    ARX &  ARMAX &  ARIMAX &   LSTM &    RNN &  HELIOS  \\
\midrule
$\text{R}^2$  & 0.59 &     0.76 &  0.59 &  0.58 &   0.56 &  0.78 &  0.64 &  0.71 &  0.36 &  0.35 &  0.40 &  0.76 &  0.47 &  \bfn{0.87}\\
RMSE  & 194.82 &    73.32 &  194.72 &  97.91 &  100.42 &  70.49 &  92.41 &  83.32 &  123.52 &  124.57 &  119.39 &  75.37 &  112.03 & \bfn{ 52.36}\\
MAE  & 153.06 &    40.53 &  152.98 &  72.13 &  74.23 &  40.22 &  53.02 &  45.68 &  83.74 &  79.81 &  81.88  &  38.41 &  70.09 & \bfn{25.88}  \\
MAPE  & 182.77 &    27.55 &  182.66 &  75.18 &  76.26 &  26.94  &  33.55 &  29.03 &  70.17 &  57.18 &  70.94 &  24.47 &  52.53 &  \bfn{20.08} \\
\bottomrule
\end{tabular}
\label{tab.performance}
\vspace{-0.2cm}
\end{table*} 
\paragraph*{Benchmark Models}\cref{tab.performance} presents the performance metrics for \changes{benchmark} models trained to predict the heat load for a prediction horizon of 12 hours. 
The HELIOS model outperforms all other models in all the performance metrics. 
HELIOS has an MAE of 25.25 [W], much lower than the 38.41 [W] from the LSTM model. 
The $\text{R}^2$ indicates that the HELIOS model accounts for 87\% of the variation in modeling the heat demand, which is higher than for all other models.
The RMSE also indicates the strong performance of the HELIOS model, followed by the XGboost, and the LSTM models. Similarly, the MAPE metric indicates the strong performance of the HELIOS model, followed by the LSTM and XGBoost models. 
We remark that these results are in line with the literature, which frequently reports LSTM and XGBoost as best-performing. 
Finally, HELIOS has another significant advantage: it is entirely decomposable and can account for the space heating, hot water, and loss profiles from a low to high-level perspective.

{
\paragraph*{Medium-to-long term predictions}In addition to the short-term heat load analysis, we further assess the performance of HELIOS against static models for medium-to-long-term forecasts (i.e., daily, weekly, monthly, and bi-annual).
This comparison is based on the premise that, for long-term predictions, dynamic changes are less pertinent, making static models a suitable benchmark. 
The results are presented in \cref{tab.performance_medium_to_long}. 
\begin{table}[!t]
\setlength{\tabcolsep}{3.6 pt}
\scriptsize
\caption{MAPE of Medium-to-long term forecasts}
\label{tab.performance_medium_to_long}
\centering
\begin{tabular}{lrrrrrrr}
\hline
Period & LR & XGBoost & RR & EN & LASSO & SVR & HELIOS \\
\hline
Daily & 53.97 & 15.39 & 53.89 & 47.75 & 48.17 & 15.63 & \bfn{14.36} \\
Weekly & 46.40 & 12.82 & 46.35 & 41.38 & 41.42 & 13.41 & \bfn{10.47} \\
Monthly & 39.87 & 11.53 & 39.85 & 33.94 & 30.16 & 12.74 & \bfn{9.60} \\
Bi-annual & 13.62 & \bfn{8.34} & 13.62 & 13.01 & 12.31 & 9.53 & 9.88 \\
\hline
\end{tabular}
\vspace{-0.25cm}
\end{table}
\changes{HELIOS outperforms benchmark models in terms of MAPE across daily, weekly, and monthly periods, with similar trends observed in all other performance metrics (RMSE, $\text{R}^2$, MAE).} 
These results confirm the significance of incorporating precise contextual information to improve forecast accuracy in the short to medium term. 
However, in the bi-annual forecasts, XGBoost slightly surpasses HELIOS, suggesting that traditional models can perform better in longer-term forecasts (under the drawback of a lack of explainability, as discussed in more detail in the following paragraph).
}
\vspace{-.5cm}

%---------------------------------------------------------------------------
%              New Subsection
%---------------------------------------------------------------------------
{
\subsection{Discussion}
\label{ss.discussion}
The HELIOS model can disaggregate the delivered heat load into space heating, domestic hot water, and heating piping loss (as can be seen in \cref{fig.fig_forecasting_january2019,fig.fig_forecasting_august2019}, where the profiles for winter and summer are displayed). From these results, space heating clearly predominates in the winter, whereas hot water consumption is more significant during summer; this is the expected behavior. Utilizing this information, the HELIOS model enables its validation by examining each disaggregated component and their anticipated seasonal behaviors.
Hence, HELIOS provides explainable results, enabling the model understanding reflecting the expected energy dynamics.

It is noteworthy to mention that contextual information is a key component in the HELIOS model. Tailoring HELIOS to represent diverse building characteristics and usage patterns accurately necessitates an in-depth understanding of its mechanisms and fine adjustments of priors and contexts; all the underlying equations of physics-based models remain the same. 
In practice, this can be a difficult task that requires in-depth expert knowledge.
However, the absence of contexts does not preclude the model's use, albeit with reduced accuracy, as evidenced in our evaluations discussed in \cref{ss.evaluating_helios}; for example, the HELIOS-NC performs comparably to LSTM and XGBoost. Moreover, while our results demonstrate HELIOS's application to residential buildings, it can be extended to other building types by defining appropriate contexts. For example, applying HELIOS to buildings with different heat patterns, such as public buildings with distinct operating hours or hospitals with continuous operation, may require (not mandatory) adjustments in the model's priors and contexts. Therefore, HELIOS can be suitable for large-scale applications, even when not all contexts for all buildings are available. 

}

%---------------------------------------------------------------------------
%              New Section
%---------------------------------------------------------------------------

\vspace{-0.2cm}
\section{Conclusions}
\label{sec.conclusion}
\changes{In this paper, we have presented HELIOS, an AI model for modeling and predicting the heat load in district heating systems (DHS).} By incorporating both physical and expert knowledge in the learning stage and integrating the functional form and prior information into the model structure, HELIOS outperforms all considered \changes{benchmark} methods. {Contextual information is a key component in increasing the model performance and plays an important (but not mandatory) element of our model}.

The inherent explainability and interpretability of HELIOS provide valuable insights for various applications, e.g., to identify factors contributing to the overall head load. 
In contrast to black-box models such as XGBoost and LSTM, HELIOS offers transparency and a clear understanding of the individual components driving the heat load. This interpretability of the model fosters a deeper understanding of the results and facilitates informed decision-making based on predictions.
Overall, the accuracy, explainability, and disaggregation capabilities of HELIOS make it a powerful tool for stakeholders to thoroughly understand heat load patterns and implement effective strategies for efficient heat load management.

{
Our current contributions on developing the HELIOS model framework itself; thus we focused on comparisons with general statistical methods.
In future work, we aim to compare the HELIOS model against more specialized prediction models for DHS, apply the HELIOS to different building types and apply HELIOS in a DHS connected a large number of buildings.
}

\vspace{-.25cm}

% % % ======================================================================
%\section{References}
%\bibliographystyle{abbrv}
%---------------------------------------------------------------------------
%              Bibliography
%---------------------------------------------------------------------------
\bibliographystyle{IEEEtran}

\begin{thebibliography}{10}
  \providecommand{\url}[1]{#1}
  \csname url@samestyle\endcsname
  \providecommand{\newblock}{\relax}
  \providecommand{\bibinfo}[2]{#2}
  \providecommand{\BIBentrySTDinterwordspacing}{\spaceskip=0pt\relax}
  \providecommand{\BIBentryALTinterwordstretchfactor}{4}
  \providecommand{\BIBentryALTinterwordspacing}{\spaceskip=\fontdimen2\font plus
  \BIBentryALTinterwordstretchfactor\fontdimen3\font minus \fontdimen4\font\relax}
  \providecommand{\BIBforeignlanguage}[2]{{%
  \expandafter\ifx\csname l@#1\endcsname\relax
  \typeout{** WARNING: IEEEtran.bst: No hyphenation pattern has been}%
  \typeout{** loaded for the language `#1'. Using the pattern for}%
  \typeout{** the default language instead.}%
  \else
  \language=\csname l@#1\endcsname
  \fi
  #2}}
  \providecommand{\BIBdecl}{\relax}
  \BIBdecl
  
  \bibitem{benonysson1995operational}
  A.~Benonysson, B.~B{\o}hm, and H.~F. Ravn, ``Operational optimization in a district heating system,'' \emph{Energy conversion and management}, vol.~36, no.~5, pp. 297--314, 1995.
  
  \bibitem{Mbiydzenyuy2022}
  G.~Mbiydzenyuy, S.~Nowaczyk, H.~Knutsson, D.~Vanhoudt, J.~Brage, and E.~Calikus, ``Opportunities for machine learning in district heating,'' \emph{Applied Sciences}, vol.~11, no.~13, 2021.
  
  \bibitem{Zdravkovic2022}
  M.~Zdravković, I.~Ćirić, and M.~Ignjatović, ``Explainable heat demand forecasting for the novel control strategies of district heating systems,'' \emph{Annual Reviews in Control}, vol.~53, pp. 405--413, 2022.
  
  \bibitem{Nielsen2006}
  H.~A. Nielsen and H.~Madsen, ``Modelling the heat consumption in district heating systems using a grey-box approach,'' \emph{Energy and Buildings}, vol.~38, no.~1, pp. 63--71, 2006.
  
  \bibitem{Lund2020}
  P.~D. Lund and V.~Arabzadeh, ``Modelling city-scale transient district heat demand by combining physical and data-driven approach,'' \emph{Applied Thermal Engineering}, vol. 178, p. 115590, 2020.
  
  \bibitem{Willcox2021}
  K.~E. Willcox, O.~Ghattas, and P.~Heimbach, ``The imperative of physics-based modeling and inverse theory in computational science,'' \emph{Nature Computational Science}, vol.~1, no.~3, pp. 166--168, 2021.
  
  \bibitem{aoun2019modelling}
  N.~Aoun, R.~Bavi{\`e}re, M.~Vall{\'e}e, A.~Aurousseau, and G.~Sandou, ``Modelling and flexible predictive control of buildings space-heating demand in district heating systems,'' \emph{Energy}, vol. 188, p. 116042, 2019.
  
  \bibitem{verrilli2016model}
  F.~Verrilli, S.~Srinivasan, G.~Gambino, M.~Canelli, M.~Himanka, C.~Del~Vecchio, M.~Sasso, and L.~Glielmo, ``Model predictive control-based optimal operations of district heating system with thermal energy storage and flexible loads,'' \emph{IEEE Transactions on Automation Science and Engineering}, vol.~14, no.~2, pp. 547--557, 2016.
  
  \bibitem{Dotzauer2002}
  E.~Dotzauer, ``Simple model for prediction of loads in district-heating systems,'' \emph{Applied Energy}, vol.~73, no.~3, pp. 277--284, 2002.
  
  \bibitem{Potocnik2015}
  P.~Potocnik, E.~Strmcnik, and E.~Govekar, ``Linear and neural network-based models for short-term heat load forecasting,'' \emph{Journal of Mechanical Engineering}, vol.~61, pp. 543--550, 2015.
  
  \bibitem{Idowu2016}
  S.~Idowu, S.~Saguna, C.~Åhlund, and O.~Schelén, ``Applied machine learning: Forecasting heat load in district heating system,'' \emph{Energy and Buildings}, vol. 133, pp. 478--488, 2016.
  
  \bibitem{Dahl2017}
  M.~Dahl, A.~Brun, and G.~B. Andresen, ``Using ensemble weather predictions in district heating operation and load forecasting,'' \emph{Applied Energy}, vol. 193, pp. 455--465, 2017.
  
  \bibitem{Suryanarayana2018}
  G.~Suryanarayana, J.~Lago, D.~Geysen, P.~Aleksiejuk, and C.~Johansson, ``Thermal load forecasting in district heating networks using deep learning and advanced feature selection methods,'' \emph{Energy}, vol. 157, pp. 141--149, 2018.
  
  \bibitem{Xue2019}
  P.~Xue, Y.~Jiang, Z.~Zhou, X.~Chen, X.~Fang, and J.~Liu, ``Multi-step ahead forecasting of heat load in district heating systems using machine learning algorithms,'' \emph{Energy}, vol. 188, p. 116085, 2019.
  
  \bibitem{Bunning2020}
  F.~Bünning, P.~Heer, R.~S. Smith, and J.~Lygeros, ``Improved day ahead heating demand forecasting by online correction methods,'' \emph{Energy and Buildings}, vol. 211, p. 109821, 2020.
  
  \bibitem{Bergsteinsson2021}
  H.~G. Bergsteinsson, J.~K. Møller, P.~Nystrup, Ólafur Pétur~Pálsson, D.~Guericke, and H.~Madsen, ``Heat load forecasting using adaptive temporal hierarchies,'' \emph{Applied Energy}, vol. 292, p. 116872, 2021.
  
  \bibitem{Castellini2021}
  A.~Castellini, F.~Bianchi, and A.~Farinelli, ``Predictive model generation for load forecasting in district heating networks,'' \emph{IEEE Intelligent Systems}, vol.~36, no.~4, pp. 86--95, 2021.
  
  \bibitem{Kato2008}
  K.~Kato, M.~Sakawa, K.~Ishimaru, S.~Ushiro, and T.~Shibano, ``Heat load prediction through recurrent neural network in district heating and cooling systems,'' in \emph{{SMC}}.\hskip 1em plus 0.5em minus 0.4em\relax {IEEE}, 2008, pp. 1401--1406.
  
  \bibitem{Lin2020}
  T.~Lin, Y.~Pan, G.~Xue, J.~Song, and C.~Qi, ``A novel hybrid spatial-temporal attention-lstm model for heat load prediction,'' \emph{IEEE Access}, vol.~8, pp. 159\,182--159\,195, 2020.
  
  \bibitem{Song2020}
  J.~Song, G.~Xue, X.~Pan, Y.~Ma, and H.~Li, ``Hourly heat load prediction model based on temporal convolutional neural network,'' \emph{IEEE Access}, vol.~8, pp. 16\,726--16\,741, 2020.
  
  \bibitem{Lu2021}
  Y.~Lu, Z.~Tian, R.~Zhou, and W.~Liu, ``Multi-step-ahead prediction of thermal load in regional energy system using deep learning method,'' \emph{Energy and Buildings}, vol. 233, p. 110658, 2021.
  
  \bibitem{Ntakolia2022}
  C.~Ntakolia, A.~Anagnostis, S.~Moustakidis, and N.~Karcanias, ``Machine learning applied on the district heating and cooling sector: a review,'' \emph{Energy Systems}, vol.~13, pp. 1--30, 2022.
  
  \bibitem{Rudin2022}
  C.~Rudin, ``Why black box machine learning should be avoided for high-stakes decisions, in brief,'' \emph{Nature Reviews Methods Primers}, vol.~2, no.~81, 2022.
  
  \bibitem{Souza2022}
  F.~Souza, T.~Offermans, R.~Barendse, G.~J. Postma, and J.~J. Jansen, ``Contextual mixture of experts: Integrating knowledge into predictive modeling,'' \emph{{IEEE} Trans. Ind. Informatics}, vol.~19, no.~8, pp. 9048--9059, 2023.
  
  \bibitem{Teshima2021}
  T.~Teshima and M.~Sugiyama, ``Incorporating causal graphical prior knowledge into predictive modeling via simple data augmentation,'' in \emph{{UAI}}, ser. Proceedings of Machine Learning Research, vol. 161.\hskip 1em plus 0.5em minus 0.4em\relax {AUAI} Press, 2021, pp. 86--96.
  
  \bibitem{Frederiksen2013}
  S.~Frederiksen and S.~Werner, \emph{District Heating and Cooling}.\hskip 1em plus 0.5em minus 0.4em\relax Professional Publishing SVC.\, 2013.
  
  \bibitem{DUBOIS2014}
  D.~Dubois and H.~Prade, ``Possibilistic logic — an overview,'' in \emph{Computational Logic}, ser. Handbook of the History of Logic, J.~H. Siekmann, Ed.\hskip 1em plus 0.5em minus 0.4em\relax North-Holland, 2014, vol.~9, pp. 283--342.
  
  \bibitem{Santin2017}
  O.~Guerra-Santin and S.~Silvester, ``Development of dutch occupancy and heating profiles for building simulation,'' \emph{Building Research \& Information}, vol.~45, no.~4, pp. 396--413, 2017.
  
  \bibitem{Badings2020}
  V.~Rostampour, T.~S. Badings, and J.~M.~A. Scherpen, ``Demand flexibility management for buildings-to-grid integration with uncertain generation,'' \emph{Energies}, vol.~13, no.~24, 2020.
  
  \bibitem{task44_annex38}
  M.~Y. Haller, R.~Dott, J.~Ruschenburg, F.~Ochs, and J.~Bony, ``The reference framework for system simulations of the iea shc task 44 / hpp annex 38: Part a - general simulation boundary conditions,'' IEA Solar Heating and Cooling Programme, Tech. Rep., 2013.
  
  \bibitem{Banks2012}
  D.~Banks, \emph{An Introduction to Thermogeology: Ground Source Heating and Cooling}.\hskip 1em plus 0.5em minus 0.4em\relax Wiley, 2012.
  
  \bibitem{Calixto2021}
  S.~Calixto, M.~Cozzini, and G.~Manzolini, ``Modelling of an existing neutral temperature district heating network: Detailed and approximate approaches,'' \emph{Energies}, vol.~14, no.~2, 2021.
  
  \end{thebibliography}

%\section*{References}
% References follow the acknowledgments. Use unnumbered first-level
% heading for the references. Any choice of citation style is acceptable
% as long as you are consistent. It is permissible to reduce the font
% size to \verb+small+ (9 point) when listing the references. {\bf
%   Remember that you can use a ninth page as long as it contains
%   \emph{only} cited references.}
% \medskip
% 
% 
% 
% [1] Alexander, J.A.\ \& Mozer, M.C.\ (1995) Template-based algorithms
% for connectionist rule extraction. In G.\ Tesauro, D.S.\ Touretzky and
% T.K.\ Leen (eds.), {\it Advances in Neural Information Processing
%   Systems 7}, pp.\ 609--616. Cambridge, MA: MIT Press.
% 
% [2] Bower, J.M.\ \& Beeman, D.\ (1995) {\it The Book of GENESIS:
%   Exploring Realistic Neural Models with the GEneral NEural SImulation
%   System.}  New York: TELOS/Springer--Verlag.
% 
% [3] Hasselmo, M.E., Schnell, E.\ \& Barkai, E.\ (1995) Dynamics of
% learning and recall at excitatory recurrent synapses and cholinergic
% modulation in rat hippocampal region CA3. {\it Journal of
%   Neuroscience} {\bf 15}(7):5249-5262.

%\appendix
%\section{Top error rank plots}
%\label{s.appendixA}
%In here, the error plots for the feature selection are presented.

\end{document}